\documentclass[useAMS,usenatbib,letterpaper]{mn2e}
\usepackage[totalwidth=480pt,totalheight=680pt,layoutvoffset=0.5cm]{geometry}

\usepackage[T1]{fontenc}
\usepackage{aecompl}

 \usepackage{graphicx,subfig}
 \usepackage{amssymb}
 \usepackage{booktabs}
\title[Transit photometry of nearby debris discs]{Feasibility of transit photometry of nearby debris discs}

\author[Zeegers, Kenworthy and Kalas]{
S.T. Zeegers$^{1}$$^{,3}$\thanks{E-mail: zeegers@strw.leidenuniv.nl},
M.A. Kenworthy$^{1}$, P. Kalas$^{2}$\\
$^{1}$ Leiden Observatory, Leiden University, P.O. Box 9513, 2300 RA
Leiden, the Netherlands\\
$^{2}$ Astronomy Department, University of California, Berkeley, CA 94720\\
$^{3}$ SRON-Netherlands Institute for Space Research, Sorbonnelaan 2, 3584 CA, Utrecht, The Netherlands}

\begin{document}

\date{Accepted for publication 20 December 2013}

\maketitle

\label{firstpage}

\begin{abstract}
Dust in debris discs is constantly replenished by collisions between larger objects. 
In this paper, we investigate a method to detect these collisions. 
We generate models based on recent results on the Fomalhaut debris disc, 
where we simulate a background star transiting behind the disc, due to the proper motion of Fomalhaut. 
By simulating the expanding dust clouds caused by the collisions in the debris disc, 
we investigate whether it is possible to observe changes in the brightness of the background star. 
We conclude that in the case of the Fomalhaut debris disc, changes in the optical depth can be observed, 
with values of the optical depth ranging
from $10^{-0.5}$ for the densest dust clouds to $10^{-8}$ for the most
diffuse clouds with respect to the background optical depth
of $\sim1.2\times10^{-3}$. 
\end{abstract}


\section{Introduction}
Debris discs are circumstellar belts of dust and debris around stars. These discs are analogous to the Kuiper belt 
and the asteroid belt in our own Solar system.  
They provide a stepping stone in the study of planet formation, because the evolution of a star's
debris disc is indicative of the evolution of its planetesimal belts 
\citep{/disks/strw1/zeegers/major_project/papers/Wyatt08}. 

Debris discs can be found around both young and more evolved stars. For the youngest stars, the dust in the disc can be considered a remnant of the protoplanetary disc. 
In these young debris discs, gas might still be present, which is not the case for debris discs around more evolved (main sequence) stars. For older stars, 
it is more likely that debris discs indicate the place where a planet has failed to form. This can either be because the formation time-scale was too long, 
like in the Solar system's Kuiper Belt, or because the debris disc was stirred up by gravitational interactions of other planets before a planet could form, 
which probably happened in the Solar system's Asteroid belt 
\citep{/disks/strw1/zeegers/major_project/papers/Wyatt08}. 
The dust in debris discs is thought to be constantly replenished by collisions between the planetesimals. 
These planetesimals start to grow in the disc during the protoplanetary disc phase. 
Models indicate that when the planetesimals begin to reach the size of $\sim2000$ km in
diameter, the process of growth reverses and the disc begins to erode. The dynamical perturbations of these large objects 
(with diameters $>$ 2000 km) stir up the disc and start a cascade of collisions 
~\citep{/disks/strw1/zeegers/major_project/papers/Kenyon05}.
However, it is not clear from these models how the dust is replenished over a time-scale of more than 100 Myr 
\citep{/disks/strw1/zeegers/major_project/papers/Wyatt08}.

The first debris discs were discovered with the Infrared Astronomical Satellite (\emph{IRAS};
\citep{/disks/strw1/zeegers/major_project/papers/Neugebauer84}), 
which measured the excess emission in the infrared caused by dust in the debris disc. The dust is heated by the central star
and therefore re-emits thermal radiation, which causes the observed spectrum of the system to deviate from that of a stellar
black body radiation curve. 
The debris disc around Vega was the first debris disc discovered in this way 
\citep{/disks/strw1/zeegers/major_project/papers/Aumann84}
and after that more than 100 discs have been subsequently discovered. 
Observations from recent surveys indicate that at least 15 per cent of FGK stars and 32 per cent of A stars have a detectable 
amount of circumstellar debris (\citet{2014MNRAS.437.3288B}, \citet{/disks/strw1/zeegers/major_project/papers/Bryden06}, \citet{2007ApJ...658.1312M},
\citet{2008ApJ...677..630H},\citet{2009MNRAS.397..757G} and \citet{2006ApJ...653..675S}).

Until improved coronagraph techniques became available, the only ground-based resolved example of a debris disc 
observed in scattered light at an optical wavelength of $0.89\,\mu$m was the debris disc
of Beta Pictoris \citep{/disks/strw1/zeegers/major_project/papers/Smith84}. 
However, during the past decade many resolved debris discs have been observed at optical and near-infrared
wavelengths. Debris discs have been observed in scattered light using the Advanced Camera for Surveys (ACS; 
\citep{2004IAUS..221..449C})
as well as the Space Telescope Imaging Spectrograph (STIS) 
and in the near-infrared ($1.1\,\mu$m) using the Near-Infrared Camera and Multi-Object Spectrometer 
(NICMOS)  
combined with the usage of coronagraphs on the \emph{Hubble Space Telescope (HST)}. 
Examples of debris disc observed with these instruments are: the debris disc of 
HD 202628 observed with STIS (see Figure~\ref{fig:HD 202628}
and \citet{/disks/strw1/zeegers/major_project/papers/Krist12}), the debris disc of AU Microscopii
\citep{/disks/strw1/zeegers/major_project/papers/Krist05} with the ACS and the NICMOS image of the debris disc HR 4769A
\citep{1999ApJ...513L.127S}.
Debris discs are also observed at infrared and (sub)millimetre wavelengths
where the dust emits the reprocessed stellar light as thermal radiation, for example the debris disc of Epsilon Eri
observed at $850\,\mu$m with the Submillimetre Common-User Bolometer Array (SCUBA) at the James Clerk Maxwell Telescope
\citep{1998ApJ...506L.133G}
and Beta Pictoris observed with Herschel Photodetector Array Camera \& Spectrometer (PACS) and Spectral and Photometric
Imaging Receiver (SPIRE; \citep{2010A&A...518L.133V}).  
These direct observations of debris discs show a wide variety of disc morphology. Some of these discs have narrow
dust rings, while other discs are more widespread. Systems can have multiple and even warped discs. 
Models show that the shape of debris discs can be caused by shepherding planets around these rings 
(\citet{/disks/strw1/zeegers/major_project/papers/Deller05};
\citet{/disks/strw1/zeegers/major_project/papers/Quillen07}). The first hints that this might be the case are given by
observations of $\beta$ Pictoris (\citet{/disks/strw1/zeegers/major_project/papers/Lagrange10};
\citet{/disks/strw1/zeegers/major_project/papers/Quanz10}) and HD 100456 
\citep{/disks/strw1/zeegers/major_project/papers/Quanz13}.

One such a star with a debris disc is the nearby [$7.668 \pm 0.03$ pc \citep{/disks/strw1/zeegers/major_project/papers/Perryman97}] 
A star Fomalhaut.
The most prominent feature of the disc is the main dust ring at a radius of 140 au from the star 
\citep{/disks/strw1/zeegers/major_project/papers/Kalas05},
which is $\sim25$ au wide. This debris disc has been imaged by the 
\emph{HST} in 2005 \citep{/disks/strw1/zeegers/major_project/papers/Kalas05} and more recently by 
the Herschel Space Telescope \citep{/disks/strw1/zeegers/major_project/papers/Acke12_updated} and the Atacama Large Milimeter Array 
(ALMA) \citep{/disks/strw1/zeegers/major_project/papers/Boley12}.
The dust around the Fomalhaut debris disc has been observed in both reflected optical light and at $10 -100 \mu \mathrm{m}$ wavelength,
where the thermal radiation of the dust in the disc is observed
radiation \citep{/disks/strw1/zeegers/major_project/papers/Holland98}.

The ring has a mass-loss rate of $2\times10^{21} \,\mathrm{g\,yr}^{-1}$ \citep{/disks/strw1/zeegers/major_project/papers/Acke12_updated}, 
which can be compared to the loss of the total mass of 
the rings of Saturn per year. This huge amount of mass suggests a high collision rate.
\citet{/disks/strw1/zeegers/major_project/papers/Wyatt02} investigated the possibility of large dust clumps in the 
Fomalhaut debris disc due to collisions between large planetesimals ($>1400 \,\mathrm{km}$) in order to 
explain a residual arc of 450$\,\mu$m emissions approximately 100 au from the star. These collisions would make
an observational detectable clumpy morphology. Such a dust clump may be detected in the debris disc of Beta Pictoris.
\citet{1995A&A...299..557L} conclude in their paper that the brightness variation in this
star on a time-scale from 1979 until 1982 can be attributed to either occultation of the star by a clumpy dust cloud or a planet.  
However, recent observations of the Fomalhaut debris disc in thermal emission show a smooth structure 
to the debris,
which hints at a high dust replenishment rate by numerous collisions 
\citep{/disks/strw1/zeegers/major_project/papers/Acke12_updated}.
The colliding planetesimals will have diameters smaller than 100 km 
(\citet{/disks/strw1/zeegers/major_project/papers/Wyatt02}; 
\citet{/disks/strw1/zeegers/major_project/papers/Greaves04};
\citet{/disks/strw1/zeegers/major_project/papers/Quillen07}) and therefore the dust clouds will be difficult to 
detect either in reflected optical light or at longer wavelengths.

Current observations of debris discs show us the distribution of small dust particles, with radii from $10^{-5}$ m to 
0.2 m \citep{/disks/strw1/zeegers/major_project/papers/Wyatt02}. We are not able to directly
observe planetesimals or large boulders. This means that we do not have an observational confirmation of 
the distribution of these larger parent particles. Observing the debris resulting from collisions would make it possible
to put constraints on the particle-size distribution in debris discs.
In this paper, we explore a technique that would make it possible to indirectly observe collisions between large
planetesimals. 
When a background object, like a star, passes behind a debris disc, dust generating collisions will cause the star to dim slightly. 
The change in brightness depends on the optical depth of the dust clouds, which in turn depends on the amount of debris created in the 
collision between two planetesimals. 

The outline of the paper is as follows. Section 2 explains the method we use to observe collisions in debris discs in 
more detail and gives a general introduction to the Fomalhaut debris disc.
In Section 3, we explain the theoretical background of the collision model used. Section 4 explains
the difference between three models of collisions in the debris disc of Fomalhaut. 
Section 5 shows differences in the observations for systems with different inclinations.
In Section 6 we show other debris discs with background objects passing behind the disc,
and we conclude in Section 7 with a summary and a discussion of our results.

\section{Detecting collisions in debris disc by observing a background star}

\begin{figure*}
 \begin{minipage}{140mm}
  \begin{center}  
  \includegraphics[scale=1.0]{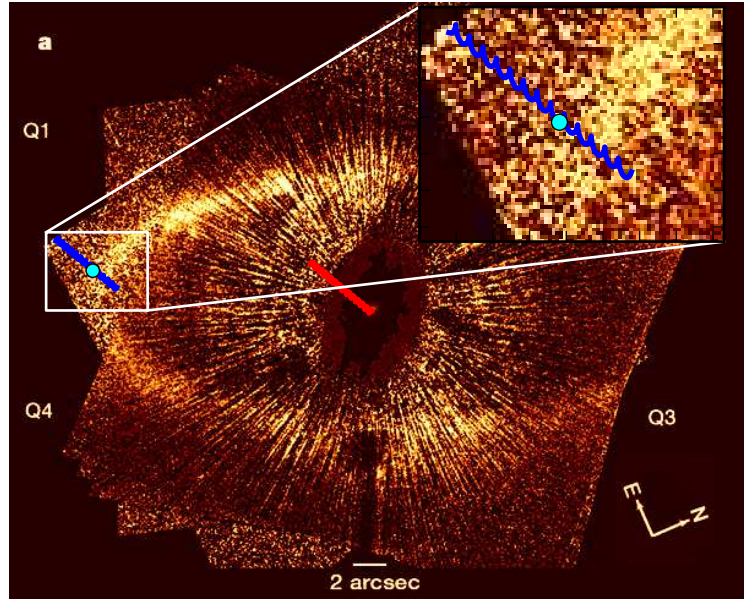}
  \caption{\small{Debris disc of Fomalhaut \citep{/disks/strw1/zeegers/major_project/papers/Kalas05} 
  with background star, that started pass behind the disc in
  January 2012 and will take 4 years to transit the disc. After 20 years the star will have a second transit. The 
  position of the star at September 15 2012 is indicated by the blue dot. 
  The joint motion of proper motion and parallax is shown by the 
  blue line.}}
  \label{fig:fomalhaut_withtransit}
    \end{center}
  \end{minipage}
\end{figure*}

In this project, we investigate whether it is possible to observe collisions between planetesimals in debris discs
by observing changes in the brightness while a distant star transits behind 
the disc. By observing changes in optical depth, we can deduce 
the size distribution of the colliding objects. In Fig.~\ref{fig:fomalhaut_withtransit} we can see that such a transiting event 
has already started in the case of the debris disc surrounding Fomalhaut. The blue dot indicates the location of 
a background star at 2012 September 15.
The star is already behind the outer part of the visible debris ring. 

If we want to observe the dust particles in the debris disc in optical light, we only observe the light reflected from 
the central star. These particles, however, 
have a low albedo and reflect only a small fraction of incident light for the Fomalhaut debris disc.
\citet{/disks/strw1/zeegers/major_project/papers/Acke12_updated} 
used a mixture of 32 per cent silicates, 10 per cent iron sulphide, 13 per cent amorphous carbon and 45 per cent water ice,
proposed by \citet{/disks/strw1/zeegers/major_project/papers/Min11}. 
This mixture is in agreement with the composition of comets which have an albedo of
3-4 per cent 
\citep{/disks/strw1/zeegers/major_project/papers/Weaver03}. 
One advantage of using a background star is that we can use the physical size of the dust to block light independent of the 
albedo, so each dust particle will block part of the light coming from this star. 
We can therefore measure the contribution of all the debris particles that originated from a collision when the star
passes behind such a cloud of debris.

Fomalhaut is not the only star with debris disc that has a transiting background object. 
A table with candidate debris discs for upcoming transits can be found in Section 6.

\subsection{Proper motion and Parallax}

When we want to follow the position of the star as it moves behind the debris disc, we must take into account that
Fomalhaut is a nearby star. At a distance of $7.668 \pm 0.03$ pc \citep{/disks/strw1/zeegers/major_project/papers/Perryman97}
it has a significant parallactic motion. 
Combined with the high proper motion of the star the joint displacement can be seen in 
Fig.~\ref{fig:fomalhaut_withtransit}, which shows the Fomalhaut debris disc, the background star
\citep{/disks/strw1/zeegers/major_project/papers/Kalas05} and their combined motion as the epicyclic motion across the 
debris disc.

The background star may be a G star with a V-band magnitude of $\sim16$ (Kalas and Kenworthy, private communication). 
With this information, we can derive the distance of the star and its effective size at Fomalhaut. 
Follow up observations are being done to determine the spectral type and to tighter constrain the 
stellar magnitude. 

\subsection{Fomalhaut debris disc}

In this paper, we will focus on the Fomalhaut debris disc, because this debris disc has a background star that will transit 
behind the disc from 2013 to 2015 and the properties of this disc are well studied. 

Fomalhaut is one of the first main-sequence stars shown to have a debris discs around it 
\citep{/disks/strw1/zeegers/major_project/papers/Aumann84}. Fomalhaut's spectral
energy distribution has an 
infrared excess above that of a model stellar chromosphere, giving already an indication of the presence of a debris disc. 
The infrared excess above that of the stellar photosphere is caused by thermal emission of the micron-sized dust particles 
in the disk. Fomalhaut is an A3V star with a mass of $1.92\pm0.02 M_{\sun}$ 
and the age of the star is estimated to be $450 \pm 40 \,\mathrm{ Myr}$ 
\citep{/disks/strw1/zeegers/major_project/papers/Mamajek12} based on 
comparison of its HR-diagram parameters to modern evolution tracks and age dating of its common proper motion companion.
A precise determination of the age
of the star is important to understand how the Fomalhaut debris disc evolved.

In the case of Fomalhaut, the star is 
almost 60 000 times brighter than its surrounding debris disc, with the disc having a V-band magnitude of 21 per 
$\mathrm{arcsec}^2$ \citep{/disks/strw1/zeegers/major_project/papers/Kalas05} 
and the star has a V-band magnitude of 1.16 
\citep{/disks/strw1/zeegers/major_project/papers/Gontcharov06}.

\subsubsection{Properties of the disc and the companion Fomalhaut b}

The disc has an inclination of $65^{\circ}.6$ from edge on. 
\citet{/disks/strw1/zeegers/major_project/papers/Kalas05} fitted an ellipse to the debris disc and found that the
centre of the belt is offset from the star by 13.4 au 
at a position angle of $340^{\circ}.5$ and the debris disc has an eccentricity of $e=0.11$. 
Assuming the star and the belt are coplanar, the projected offset is
15.3 au in the plane of the belt. 
The largest planetesimals (also called parent bodies) in the debris ring sit at the top of a 
collisional cascade \citep{/disks/strw1/zeegers/major_project/papers/Chiang09}. 
These parent bodies are believed to be in a nearly circular ring at a radius of $\sim140 \mathrm{au}$ 
from the star. 

The planet candidate Fomalhaut b can be found at the inner
side of the debris ring. Fomalhaut b was first detected by \citet{/disks/strw1/zeegers/major_project/papers/Kalas08}.
The detected point source was verified in multiple data sets and was comoving with the star except for a small 
offset between the epochs which suggested an orbital motion in counterclockwise direction. 
The detection was confirmed in an independent analysis by 
\citet{/disks/strw1/zeegers/major_project/papers/Currie12} and 
\citet{2013ApJ...769...42G}, and most recently in new 
\emph{HST} observations presented by \citet{2014IAUS..299..204K}. 
\citet{2014IAUS..299..204K} estimated the planet mass 
to be in the range between our solar system's dwarf planets and Jupiter.
Fomahaut b was not detected in ground-based observations at 1.6 and 3.8 $\mu$m 
\citep{/disks/strw1/zeegers/major_project/papers/Kalas08}. Hereby, they 
established that the brightness at 0.6$\,\mu$m originates from non-thermal sources, probably scattering of light 
by dust. This dust can originate from collisions between the planet and planetesimals in the debris ring when the
planet crosses the debris ring. According to \citet{2014IAUS..299..204K}, 
it is unlikely that Fomalhaut b would cross the disk but it 
cannot be ruled out either. 
A possible alternative explanation is that it is a super-Earth mass planet embedded in a planetesimal swarm
\citep{/disks/strw1/zeegers/major_project/papers/Kennedy11}.  

\subsubsection{Can we see planetesimals in the Fomalhaut debris disk using the background star?}

\citet{/disks/strw1/zeegers/major_project/papers/Acke12_updated} 
find that to replenish the dust in the debris disc a population of $2.6\times10^{11}$ 
10-km-sized planetesimals or $8.3\times10^{13}$ 1-km-sized planetesimals undergoing a collisional cascade is needed. 
This huge number of planetesimals with a diameter of 1 or 10 km raises the question whether we can see a planetesimal 
passing in front of a background star. With $8.3\times10^{13}$ planetesimals of 1 km in diameter, the chance of observing one of these 
planetesimals with the star is very small, namely $\sim10^{-6}$. Furthermore only 0.003 percent of the star will be 
blocked by such a planetesimal, due to the fact that the star is not a point source at 
the plane of the Fomalhaut debris disc. 
Assuming a solar-type star, the background star has an effective size at Fomalhaut of 3550 km in diameter. 
For 10 km planetesimals the chance of observing 
such a planetesimal is $\sim3\times10^{-7}$. We can therefore state that it is highly unlikely that we will be able to observe
a large planetesimal blocking the star. 
For these calculations, we made an estimate of the total area of the Fomalhaut debris disc.
We model the dust ring as an annulus formed from two concentric nested ellipses. 
To calculate the area of the dust ring, we will estimate the semi major and semi minor axis of the inner ring of the 
dust ring and likewise the semi major and semi minor axis of the outer ring. These estimates where made using the
observations of \citep{/disks/strw1/zeegers/major_project/papers/Kalas05}

\begin{itemize}
\item Estimated semi major axis $(a_1)$ outer edge: 145 au
\item Estimated semi major axis $(a_2)$ inner edge: 130 au
\item Estimated semi minor axis $(b_1)$ outer edge: 85 au
\item Estimated semi minor axis $(b_2)$ inner edge: 65 au.
\end{itemize}

The area of the dust ring is given by the area of the large ellipse $A_1$ minus the area of the small ellipse $(A_2A)$:
\begin{equation}
  A_1-A_2=\pi a_1 b_1-\pi a_2 b_2=12173 \,\mathrm{au}^2
\end{equation}  

The next question we ask ourselves is: can we observe the dust clumps resulting from the collisions between planetesimals?
\citet{/disks/strw1/zeegers/major_project/papers/Wyatt02}
suggested that dust clumps from collisions can be seen in scattered light when they cover a projected area 
larger or equal to $0.2 \mathrm{ au}^2$ (angular size of 60 mas). In this example, a runaway planetesimal is impacted multiple
times by smaller planetesimals, which launches a cloud of regolith dust from its surface. 
Due to gravity, the majority of the dust will collapse back on the planetesimal. 
These dust clumps are expected to last for half an orbital period before the fragments
of the dust cloud occupy half the ring. The colliding
planetesimal causing such a dust cloud must have a mass larger than $0.01 M_{\earth}$ 
(or with a diameter of $\sim10^7\,\mathrm{m}$) in order to be seen by the \emph{HST}. 
Looking at the observations made with the Herschel space telescope 
\citet{/disks/strw1/zeegers/major_project/papers/Acke12_updated} conclude that
Fomalhaut's disc is too smooth to contain such large dust clumps. 
In the next section, we will investigate whether it is possible to observe smaller dust clumps using
changes in the brightness of the background star.

\section{Modelling Collisions in Debris Discs}

When modelling collisions between kilometre-sized objects in space, we do not have many examples of such collisions in our 
own Solar system to compare to these models. The best direct observation is the collision between 
comet Temple 1 and a component of NASA's deep impact probe \citep{/disks/strw1/zeegers/major_project/papers/Hearn05}. 
The resulting dust cloud from the collision was visible in 
scattered light for a week and the expansion rate of the dust cloud was measured to be $200\,m\,s^{-1}$
\citep{/disks/strw1/zeegers/major_project/papers/Rengel09}.
These lack of observations are the main reason for relying on models.
In this section, we discuss the most common used models and we explain the assumptions we make for our models. 
We show how we use these models to
simulate the expanding dust clouds resulting from the collisions between planetesimals and how we can apply them to the 
Fomalhaut debris disc.

\subsection{Collisional cascades}

There are many numerical models that describe collisions between planetesimals. 
Most of them are based upon the 
model of \citet{/disks/strw1/zeegers/major_project/papers/Dohnanyi69}.
This model gives a size distribution of the remnant particles based upon a collisional 
cascade. This cascade of collisions starts when the planetesimals in the belt are dynamically stirred and have 
attained such high relative velocities that the collisions become destructive 
\citep{/disks/strw1/zeegers/major_project/papers/Wyatt08}.
The equilibrium size distribution resulting from a collisional cascade is given by equatio~\ref{eq:collisionalcascade}:

\begin{equation}
 n(D) = K D^{2-3q}.
 \label{eq:collisionalcascade}
\end{equation}

In this equation, $K$ is a scaling factor and $n(D)$ is the number of particles with diameter between $D$ and $D+dD$.
The size distribution is a power law with an index $q$. 
Collisional size distributions depend strongly on the power-law index $q$. In the early work of Dohnanyi, this parameter 
was set by solving a differential equation describing the evolution of a system undergoing inelastic collisions 
for steady-state conditions. This results in a value $q=1.833$, which is in agreement with a fit to the 
distribution of asteroids in the Solar system's Asteroid belt \citep{/disks/strw1/zeegers/major_project/papers/Dohnanyi69}. 
This model is of course a simplification of reality, because it ignores that the strength of a 
planetesimal varies with size, which causes the slope of the distribution to change 
\citep{/disks/strw1/zeegers/major_project/papers/Wyatt02}.

The size distribution is assumed to hold from the smallest particles up to the planetesimals that feed the cascade. This 
means that though most of the mass of the cascade is in the large planetesimals, most of the cross-sectional area
is in the smallest dust particles \citep{/disks/strw1/zeegers/major_project/papers/Wyatt08}. 
This model can be used to describe the size or mass distribution of the whole debris disc and with some adaptions it can 
also be used to describe the size distribution of the debris from a single collision (see paragraph 3.2.1). 
For the whole belt and the calculation of the scaling parameter, we adopt the same value for the $q$ parameter as 
\citet{/disks/strw1/zeegers/major_project/papers/Dohnanyi69},
which is also the value that \citet{/disks/strw1/zeegers/major_project/papers/Acke12_updated} use, namely $q=1.833$. 
The particles that feed the cascade are planetesimals with sizes between 1 and 100 km 
\citep{/disks/strw1/zeegers/major_project/papers/Wyatt08}. These planetesimals mark
the top of the cascade. 
The smallest particles in the disc have sizes just above the blow-out size.
The blow-out size is the minimum size that a particle can have and remain in the disc. Smaller particles 
are blown out of the ring as soon as they are created due to the radiation pressure of the star.

The ratio of the force of the radiation pressure to that of stellar gravity is parameterized by $\beta$, where
\begin{equation}
 \beta=\frac{3L_{*}}{16\pi c GM_*\rho s},
\end{equation}

where $L_{*}$ is the luminosity of the star, $G$ is the gravitational constant, $c$ is the speed of light, $\rho$ is
the density of the dust particle and $s$ is the diameter of the particle. 
There is some variation in the density used in collisional cascade models of debris discs. 
\citet{/disks/strw1/zeegers/major_project/papers/Acke12_updated} suggest that the dust particles in the belt are less dense
and consider \textquotedblleft{fluffy aggregate}\textquotedblright particles. The density of the larger particles is considered to be the same as the value
of the density of comets in our own Solar system, such as Temple 1 with a density of 
$\rho=0.6 \,\mathrm{g}/\mathrm{cm}^3$ \citep{/disks/strw1/zeegers/major_project/papers/Hearn05}.
This low density is due to the high porosity of planetesimals.  
The average density of all the particles in this 
paper is set at $\rho=1\, \mathrm{g}/\mathrm{cm}^3$ following \citet{/disks/strw1/zeegers/major_project/papers/Chiang09}. 

Grains are unbound from their star when $\beta\gtrsim1/2$:
\begin{equation}
 s<s_{\mathrm{blow}}\approx\frac{3L_{*}}{8\pi cGM_*\rho}
\end{equation}
In the case of the fomalhaut debris disc this means that the smallest particles of the collisional cascade have sizes $>8\mu\mathrm{m}$
\citep{/disks/strw1/zeegers/major_project/papers/Chiang09}.

For individual collisions, we use a $q$ parameter that deviates slightly from the classical 1.833 value. 
The most common values for $q$ to describe the size distribution after a collision of two planetesimals are
between 1.9 and 2 \citep{/disks/strw1/zeegers/major_project/papers/Wyatt02}. 
The physical background of this deviating $q$ value originates in the porous structure of comets. After the collision,
the fragmentation continues due to the coalescence of flaws propagating through the impacted planetesimal. 
The planetesimal breaks apart more easily along the flaws and crumbles up in sequentially smaller fragments, 
and so the slope of the cascade becomes slightly steeper \citep{/disks/strw1/zeegers/major_project/papers/Wyatt02}. 
While the value of $q$ can fall anywhere between 1.6 and 2.6,
simulations have shown that for these collisions values between 1.9 and 2 are most common for individual collisions.
We will use a value for the index of 
$q=1.93$ which is in agreement with 
results from \citet{/disks/strw1/zeegers/major_project/papers/Campobagatin01}. 
Their analytical model predicted a value of 1.93, which was in 
agreement with their simulations.

We assume that the dust in the debris disc originates from a belt with colliding planetesimals. In the case of
Fomalhaut, this is a ring of planetesimals distributed around the mean radius of
140 au with a Gaussian standard deviation of 7 au in the radial direction and a Gaussian standard deviation of 5 au 
perpendicular to the ring.

\subsection{Catastrophic collisions and cratering events}

When two planetesimals collide there can be two outcomes of this collision, namely a cratering event or a catastrophic
collision. In the case of cratering, the impact energy is not large enough to destroy the 
target object. The result of a cratering collision is a crater in the impacted object, 
whereby some material is ejected, but the object is left mostly 
intact. 
In the case of a catastrophic collision both objects are destroyed by the impact. 
Both scenarios are described in the paper of \citet{/disks/strw1/zeegers/major_project/papers/Wyatt02}
of which we will give a short 
summary in this section. This paper focusses on collisions in the Fomalhaut debris disc.

The incident energy of two colliding planetesimals is given by equation~\ref{eq:energy}:

\begin{equation}
Q=0.5(D_{\mathrm{im}}/D)^3v_{\mathrm{col}}^2g.
 \label{eq:energy}
\end{equation}

In this equation, $D$ is the diameter of the planetesimal impacted by another planetesimal of size $D_{\mathrm{im}}$
and $g$ is the ratio of the densities of the two planetesimals. In this paper, we assume that the densities of the planetesimals
are the same, so $g=1$.
It is customary to characterize such impacts in terms of energy thresholds 
\citep{1999Icar..142....5B}.
The shattering threshold $Q_s$ is defined as the incident energy needed to break up the planetesimal. The largest 
remaining debris particle has at most a mass of half the mass of the original planetesimal
\citep{1999Icar..142....5B}.   
Collisions with $Q<Q_s$ will result in cratering whereby some material is ejected, but the larger planetesimal stays
largely intact. For planetesimals with $D>150\, \mathrm{m}$, the energy $Q_s$ might not be high enough to overcome the gravitational
binding energy of the planetesimal and some of the fragments may re-accumulate in a rubble pile
(\citet{2001Icar..149..198C}; \citet{/disks/strw1/zeegers/major_project/papers/Michel01}). In this case we 
need an energy of $Q>Q_D$ to create a catastrophic collision, where $D$ refers to the size of the largest remnant 
(that could be the rubble pile). Collisions between particles with $D<150\, \mathrm{m}$ for which $Q_s\approx Q_D$
are said to occur in the strength regime, while collisions between larger planetesimals occur in the gravity
regime. Most of the planetesimals considered here will have a diameter of 
$100 \,\mathrm{m} <D< 2 \,\mathrm{km}$,
so most of the particles will fall in the transition between the gravity and the strength regime where $Q_s\approx Q_D$.
There are several studies of how $Q_D$ and $Q_s$ vary with planetesimal size, composition (e.g. ices and rock), 
structure and other parameters such as different relative velocities and impact parameters.
For small particles, results from laboratory experiments can be used to determine these threshold values 
(\citet{1989aste.conf..240F}; \citet{1990Icar...83..156D}). Threshold energies of 
larger planetesimals can be modelled by detailed theoretical models 
(\citet{1986MmSAI..57...65H}, \citet{1990Icar...84..226H}; \citet{1994P&SS...42.1067H}) 
or by models based on the interpretation of the distribution of asteroid families
(\citet{2010Icar..207...54J}; \citet{1999Icar..141...65T}; \citet{1999Icar..141...79C})
or computational modelling using smooth particle hydrodynamics
(\citet{1999Icar..142....5B}, \citet{/disks/strw1/zeegers/major_project/papers/Love96}). 

To determine $Q$, we need to know the collision velocity ($v_{\mathrm{col}}$), which can be calculated for each
collision by determining the relative velocity and the escape velocity of the particles,

\begin{equation}
 v_{\mathrm{rel}}=f(e,I)v_k,
\end{equation}

where $v_k$ is the Keplerian velocity of the planetesimals and $f(e,I)$ is a function of the average eccentricities ($e$)
and inclinations ($I$) of the planetesimals given as $\sqrt{\frac{5}{4}e^2+I^2}$ 
(\citet{/disks/strw1/zeegers/major_project/papers/Wyatt02}
; \citet{/disks/strw1/zeegers/major_project/papers/Lissauer93}; \citet{1993Icar..106..190W}). 
At a distance of 140 au $v_k=3.6 \,\mathrm{km}\,\mathrm{s}^{-1}$
and the average orbital period of the planetesimals is 1150 yr. 
In the Fomalhaut model $f(e,I)\approx0.11$ thus, $v_{\mathrm{rel}}\approx 0.4 \,\mathrm{km}\, \mathrm{ s}^{-1}$ at a mean
distance of 140 au. The collision velocity can then be given by equation~\ref{eq:collisional_velocity} and the escape
velocity is given by equation~\ref{eq:escape_velocity} where a 
planetesimal of size $D$ is impacted by another planetesimal of size $D_{\mathrm{im}}$.

\begin{equation}
 v_{\mathrm{col}}^2=v_{\mathrm{rel}}^2+v_{\mathrm{esc}}^2(D,D_{\mathrm{im}})
 \label{eq:collisional_velocity}
\end{equation}

\begin{equation}
 v_{\mathrm{esc}}=(2/3)\pi G\rho \frac{D^3+D_{\mathrm{im}}^3}{D+D_{\mathrm{im}}}
 \label{eq:escape_velocity}
\end{equation}

For planetesimals with a diameter of 700 km (0.6 per cent of a lunar mass) or larger, the increase in impact velocity due
to gravity becomes important. After the impact of objects that find themselves in the gravity regime it is likely that the 
debris from such a collision reaccumulates into a rubble pile 
\citep{/disks/strw1/zeegers/major_project/papers/Campobagatin01}.
Since we do not consider such large objects (since the chance that they will participate in 
a catastrophic collision is too small), we do not take gravitational focusing into account. 
\citet{/disks/strw1/zeegers/major_project/papers/Wyatt02} 
show that the weakest planetesimals have sizes between 10 m and 1 km.
They calculated the energy threshold versus the diameter of the planetesimals for three different models
consisting of ice, weak ice and basalt. The threshold of the energy needed to shatter the planetesimal 
decreases for increasing size, as a result of the decreasing shattering strength of larger planetesimals. 
Planetesimals with diameters larger than 1 km have increasingly higher threshold energies in the gravity regime due to 
the extra energy required to overcome the planetesimal's gravity. 
To avoid the calculations involved in determining whether a planetesimal is shattered or not, we will assume
that all the colliding planetesimals will have a value of $Q$ high enough (i.e. above the threshold energy) 
to form a cloud of debris.
 
\subsubsection{The largest remnant}

The largest remnant is the largest particle that remains after a catastrophic collision. It is typically half of the
mass of the original planetesimal or less. In our simulations, we will assume that the mass of the largest 
remnant is always half of the planetesimal mass. 
We will assume 
that the size distribution follows a cascade model with $q=1.93$ for
fragments smaller than the second largest remnant. While the value of $q$ can fall anywhere between 1.6 and 2.6,
simulations have shown that for these collisions values between 1.9 and 2 are most common. 

The second largest remnant is given by equation~\ref{eq:largest_remnant}. In this equation, $q_c=1.93$ and $D_2$ is 
the size of the second largest remnant, which is the largest particle following the size distribution of the collisional 
cascade. All other debris particles will be smaller.  

\begin{equation}
 D_2/D=\left[\left(\frac{2-q_c}{q_c}\right)(1-f_{lr})\right]^{1/3}
 \label{eq:largest_remnant}
\end{equation}

\subsection{Collision rates}

We assume that collisions take place in a ring in the debris disc. 
Collisions can happen everywhere in this ring with a normal distribution around the central part of this ring.
These collisions produce the dust that is responsible for the observed reflected light in Fig.~\ref{fig:fomalhaut_withtransit}. 
Observations show that there is a high number of small dust particles in the Fomalhaut debris disc
with sizes below the blow-out size. 
\citet{/disks/strw1/zeegers/major_project/papers/Acke12_updated} find in their best fitting model a total mass 
of $3\times10^{24}\, \mathrm{g}$ for grains with sizes smaller than $13 \,\mu \mathrm{m}$.

The amount of dust escaping the system must be replenished by collisions at a constant rate. 
\citet{/disks/strw1/zeegers/major_project/papers/Acke12_updated} 
calculate that to keep the ring dusty enough, one needs at least a mass-loss rate of
$2\times10^{21} \,\mathrm{g}\,\mathrm{yr}^{-1}$. 
This mass-loss rate can be compared to 1000 collisions of 1 km in diameter sized planetesimals 
per day or 1 collision between 10-km-sized planetesimals per day. 
This number of planetesimals is not unreasonable since the Solar Symstem's Oort cloud is 
considered to contain a number of $10^{12}-10^{13}$ planetesimals
\citep{Weissman}.
The total mass of the Fomalhaut belt necessary to keep this collision rate stable is 110 Earth masses
\citep{/disks/strw1/zeegers/major_project/papers/Acke12_updated}.

We only treat catastrophic collisions between particles of the same size. We assume these
collisions happen and the number of collisions is
based on the mass-loss rate per day. The reason for this strategy is that we want to simulate many collisions per day. 
These simulations will take too much time if all the collisional equations are taken into account. We take a 
first look at the feasibility of observing planetesimals and therefore adopt a simple model. 
As for the size distribution we will consider planetesimals with a size up to 25 km, we ignore larger particles
because their collision rate is very small within the time frame of the background star crossing behind the debris ring.

\subsection{Debris velocities after the collision}
In most catastrophic and cratering collisions, there is enough energy left after the impact to impart the fragments 
with a velocity in random directions. This means that the debris from the collisions form an expanding clump of
material of which the center of mass follows the orbit of the former planetesimal. 
The velocity with which the cloud of debris expands after the collisions depends 
mostly on the parameter $f_{\mathrm{KE}}$, which is the kinetic energy imparted to the debris after the collision. 
The value of this parameter is not well known. From laboratory experiments of collisions between cm-sized objects a value of 
0.3-3 per cent of the impact kinetic energy is imparted to the largest remnant
\citep{/disks/strw1/zeegers/major_project/papers/Fujiwara80}. Studies of the asteroid families imply a
value of $f_{\mathrm{KE}}\approx0.1$ \citep{Davis89}.
Other simulations imply values of $f_{\mathrm{KE}}=0.2-0.4$ 
(e.g. \citep{/disks/strw1/zeegers/major_project/papers/Davis85}). 
For simplicity, we will assume a value of 
$f_{\mathrm{KE}}=0.1$ (following \citet{/disks/strw1/zeegers/major_project/papers/Wyatt02}), 
which is valid in both the strength and the gravity regime. 
The velocity of the debris particles after the collision have a range of velocities approximated by a power-law
function $f(v)=v^{-k}$, with a value of $k$ between 3.25 
\citep{/disks/strw1/zeegers/major_project/papers/Gault63} and 1.5 
\citep{/disks/strw1/zeegers/major_project/papers/Love96}.
Values of $k<2$ imply that most of the kinetic energy is carried away by the smallest dust particles. 
The velocity of these particles is also size dependent. Large particles tend to have lower velocities than small 
particles. The weakest planetesimals with sizes between 10 meters and 1 km will shatter in many particles and for 
this reason these planetesimals have the lowest ejection velocities. 
Very little energy is imparted to the largest remnant 
(\citep{/disks/strw1/zeegers/major_project/papers/Nakamura91}; \citep{/disks/strw1/zeegers/major_project/papers/Michel01}).
For simplicity, we assume that that the available kinetic energy after the collision is distributed among all the debris 
particles and no kinetic energy is imparted to the largest remnant (in the case that we consider a largest remnant 
particle in our simulations). This indicates that all the 
debris particles have the same velocity, except for the largest remnant that has zero velocity.

\citet{/disks/strw1/zeegers/major_project/papers/Wyatt02}
calculated the characteristic ejection velocity for debris particles in the Fomalhaut debris disc,
assuming that the kinetic energy is distributed among all the fragments except the largest remnant:

\begin{equation}
 0.5f_{\mathrm{KE}}E_{\mathrm{col}}=0.5(1-f_{\mathrm{lr}})Mv^2_{\mathrm{ej}}.
\end{equation}

From this equation, we can derive the ejection velocity: 

\begin{equation}
 v_{\mathrm{ej}}=\frac{f_{\mathrm{KE}}Q(D,D_{\mathrm{im}})/[1+(D_{\mathrm{im}}/D)^3]}{[1-f_{\mathrm{lr}}(D,D_{\mathrm{im}})]}.
\end{equation}

\noindent In this paper, we assume that $D\approx D_{\mathrm{im}}$ and $f_{\mathrm{lr}}\approx0.5$. 
For particles in the strength regime, we can then calculate the ejection velocity in the following way:

\begin{equation}
 v_{\mathrm{ej}}^2\approx f_{\mathrm{KE}}Q(\mathrm{D},\mathrm{D_{\mathrm{im}}}),
 \label{eq:ejectionvelocity}
\end{equation}

\noindent where we remind the reader that
\begin{equation}
 Q(\mathrm{D},\mathrm{D_{\mathrm{im}}})=0.5 (D/D_{\mathrm{im}})^3v_{\mathrm{col}}^2.
\end{equation}

\noindent Inserting $Q(\mathrm{D},\mathrm{D_{\mathrm{im}}})$ in equation~\ref{eq:ejectionvelocity}
gives

\begin{equation}
 v_{\mathrm{ej}}\approx \sqrt{0.5 f_{\mathrm{KE}}}v_\mathrm{col}(D,D_{\mathrm{im}}), 
\end{equation}

\noindent where $v_\mathrm{col}$ is given by equation~\ref{eq:collisional_velocity}.
When $f_{\mathrm{KE}}=0.1$ and we calculate $v_{\mathrm{col}}$ for planetesimals for which gravitational focusing is not important

\begin{equation}
 v_{\mathrm{ej}}\approx 90 \,\mathrm{m}/\mathrm{s}
\end{equation} 

In the gravity regime (planetesimals with $D>1 \mathrm{km}$), the debris particles have to overcome the gravity of the
largest remnant. They will gain a characteristic velocity once they are far enough away from the largest remnant,
which is shown in equation~\ref{eq:inf}: 

\begin{equation}
 v_\infty=\sqrt{v^2_{\mathrm{ej}}-v^2_{\mathrm{grav}}},
 \label{eq:inf}
\end{equation}

\noindent with

\begin{equation}
 v^2_{\mathrm{grav}}=0.4\pi G\rho D^2 [1-f_{\mathrm{lr}}(D,D_{\mathrm{im}})^{5/3}][1-f_{\mathrm{lr}}(D,D_{\mathrm{im}})]^{-1}
\end{equation}

and again we assume that no kinetic energy is imparted to the largest remnant.
The diameter of the largest planetesimals that collide in our model is 25 km. We calculate $v_\infty$ for a collision between
two such planetesimals and the result is that the expansion velocity of the dust cloud (far enough from the largest remnant)
is $v_\infty=87.6 \,\mathrm{m}\,\mathrm{s}^{-1}$. Therefore, we will assume an expansion velocity of the dust cloud of $90 \,\mathrm{m}/\mathrm{s}$ 
for every collision. If we include planetesimals with diameters of $D>100\,\mathrm{km}$ then 
$v_{\mathrm{grav}} > v_{\mathrm{ej}}$, which means that this method cannot be used for planetesimals larger than 
100 km. 
It is unlikely that planetesimals with diameters larger than 700 km are involved in catastrophic collisions and even collisions 
between 100 km planetesimals would be extremely rare \citep{/disks/strw1/zeegers/major_project/papers/Wyatt02}. 

\subsection{Extinction of light by a slab of dust particles} 

The light coming from the background star can be partially blocked by dust clouds. 
The change in brightness depends on the change in optical depth due to the particles resulting from the
collision, as can be seen in equation~\ref{eq:optical_depth}.

\begin{equation}
 I=I_0 e^{-\tau}
 \label{eq:optical_depth}
\end{equation}

The optical depth ($\tau$) can be calculated by first considering the area of all the particles that originate from the
collision. When observing at a wavelength of 0.5 $\mu\mathrm{m}$, geometric optics are relevant.  
This means that 
we don not have to consider Mie scattering or Rayleigh scattering, because the smallest dust particles (dust particles of 
$\sim 10\,\mu\mathrm{m}$) are larger than the wavelength: $D_{\mathrm{dust}} >> \lambda$ (where $\lambda$ is the wavelength and
$D_{\mathrm{dust}}$ the diameter of the smallest dust particles). 
To calculate the total optical depth of the slab of dust particles, we have to determine the extinction parameter 
of the dust particles. All the energy incident on the particle is absorbed and in addition an equal amount of 
energy is scattered (diffracted) by the particle \citep{1983asls.book.....B}. 
We assume that the dust is evenly distributed in all directions except for particles that are
smaller than the blowout-size, they are instantly removed from the cloud due to the radiation pressure. This means 
that after the collision all the particles have a velocity in random directions, which causes the dust cloud to expand 
immediately after the impact. 
During the expansion of the cloud the optical depth reduces slowly until the cloud blends in with the local background.
The optical depth is then given by equation~\ref{eq:expanding_cloud}.

\begin{equation}
 \tau=2\frac{N_{\mathrm{total}}A}{A_{\mathrm{sphere}}}
 \label{eq:expanding_cloud}
\end{equation}

In equation~\ref{eq:expanding_cloud}, $N_{\mathrm{total}}$ is the total number of particles in the dust cloud and $A$ is 
the geometric cross sectional area of a particle, so $N_{\mathrm{total}}\cdot A$ is the total surface of all the particles. 
When the particles resulting from the collision follow a 
size distribution given by the collisional cascade model, we have to take into account that the number of particles
will be different for each particle diameter.
Due to all the previous collisions in the debris disc,
the constant rate of collisions have caused an amount of dust that forms this background optical depth. 
The value can be determined from the observations of the Fomalhaut debris disc in scattered light 
(observed by \citet{/disks/strw1/zeegers/major_project/papers/Kalas05}). 
\citet{/disks/strw1/zeegers/major_project/papers/Chiang09}
use this data to calculate the value of the optical depth in the
disc in the radial and perpendicular direction.
 
\begin{equation}
 \frac{L_{IR}}{L_{*}}=\frac{2\pi R \times 2H \times \tau_R}{4\pi R^2}=\frac{H}{R}\tau_{R}
\end{equation}

In this equation, $\tau_R << 1$ is the radial geometric optical depth through the debris ring. 
The results of the  paper of \citet{/disks/strw1/zeegers/major_project/papers/Kalas05}
gave an aspect ratio of $H/R=0.025$. This gives a value of $\tau_R=1.8\times10^{-3}$.
The vertical optical depth (measured perpendicular to the midplane of the belt) is given by equation 
~\ref{eq:tau_perpendicular}.

\begin{equation}
 \tau_{\perp}=\tau_R\frac{2H}{\Delta R}=\frac{L_{IR}}{L_{*}}\frac{2R}{\Delta R}
 \label{eq:tau_perpendicular}
\end{equation}

This means that $\tau_{\perp}$ is independent of the $H$. From the results of \citet{/disks/strw1/zeegers/major_project/papers/Kalas05},
the values of $\tau_{\perp}$
can be determined. $\frac{2R}{\Delta R}\approx0.17$, so $\tau_{\perp}=5.4\times10^{-4}$.
We use a value for the background optical depth that is an average of the radial optical depth $\tau_R$ and 
the vertical optical depth $\tau_{\perp}$, because the star does not shine perpendicular through the debris disc as seen from our
line of sight. Therefore, the background value
of $\tau$ will be between these two limiting cases.
We therefore assume a value of $\tau=1.2\times10^{-3}$.
Collisions between planetesimals in the debris disc and especially collisions between 1-km-sized planetesimals 
can cause the optical depth to vary slightly along the disc.

\section{Simulations}

\subsection{Details about the simulations}

We simulated the collisions in the debris discs for three scenarios. The first two scenarios differ in distribution
of the particle sizes and the last one is a variation of the second scenario for which the case of a largest 
remnant particle has been included.
Table~\ref{table:models} shows an overview of the most important differences between 
the three scenarios.

\begin{table*}
 \begin{minipage}{140mm}
  \caption{Overview of the three models.}
  \begin{tabular}{l|c|c|c|c|l r}
  Scenario& Diameter of planetesimals (D) & Density of planetesimals & Largest remnant& Debris sizes &\\
  \hline\hline
  Scenario 1& 1 km & $1 \,\mathrm{g}\,\mathrm{cm}^{-3}$& No largest remnant&10 m - 8 $\mu$m &\\ \hline
  Scenario 2& 100 m - 25 km & $1 \,\mathrm{g}\,\mathrm{cm}^{-3}$ & No largest remnant &10 m - 8 $\mu$m&\\ \hline
  Scenario 3& 100 m -25 km & $1 \,\mathrm{g}\,\mathrm{cm}^{-3}$ & Largest remnant&$D_2\,^*$ - 8 $\mu$m&\\ \hline
  \end{tabular}
  \\[2pt]
  $^*\,D_2$ is the size of the second largest remnant.
  \label{table:models} 
\end{minipage}
\end{table*}

\begin{figure}
 \begin{center}
 \advance\leftskip-0.85cm
 \includegraphics[scale=0.8]{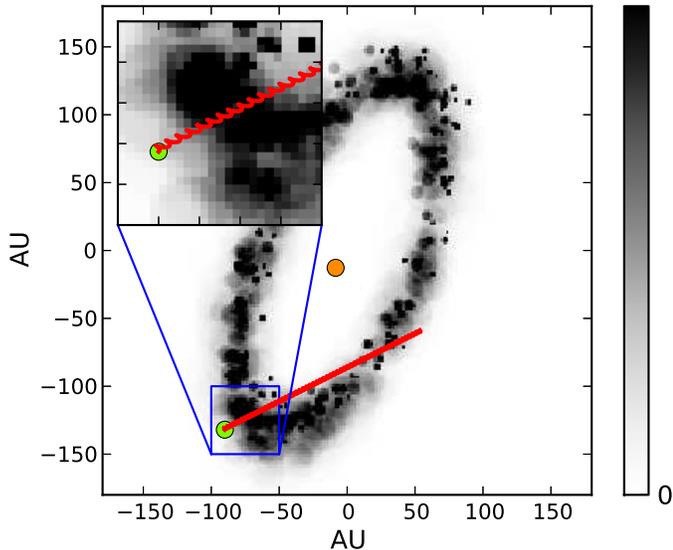}
 \caption{\small{ The figure shows a simulation of collisions in the Fomalhaut debris disc. In this simulation, we 
 exaggerated the size, Keplerian velocity and optical depth of the cloud and plot only one collision per simulated day 
 to show that the simulation is able to reproduce the dust ring
 around the star. The orange dot represents the position of Fomalhaut, the green dot shows the position of the
 background star in 2005 and the red line show the combined parallax and proper motion of the background star in the rest 
 frame of Fomalhaut. The dust ring is broader than in reality,
 due to the exaggerated expansion velocities of the dust clouds.}}
 \label{fig:fastforward}
  \end{center}
\end{figure}

\noindent All scenarios generate collisions, though the number of collisions per day in the ring differs between the models.
Our starting point for the number of collisions per unit volume is based on the result 
of \citet{/disks/strw1/zeegers/major_project/papers/Acke12_updated}, i.e. 1000 collisions in the whole disc per day. 
Using the dimensions of 
the disc \citep{/disks/strw1/zeegers/major_project/papers/Kalas08}, 
this results in a collision rate of $\sim0.004\,\mathrm{collisions}\,\mathrm{au}^{-3}\mathrm{d}^{-1}$.
Initial collisions are randomly distributed among the debris ring. To simulate this distribution, we use a Gaussian distribution
with a mean radius of 140 au from the central star and a standard deviation of $1\sigma$ of 7 au. 
The scale-height along the z-axis has a standard deviation $1\sigma$ or 5 au. 
For each collision, the position of this collision is stored. 
Every simulated day new collisions are added to the system and we keep track of the positions of the previous
collisions by calculating their displacement due to the Keplerian orbit in which all the particles find themselves. 
We only simulate collisions in a $3 \times 3 \,\mathrm{au}$ box at Fomalhaut, since the displacement of the particles 
in orbit is 0.8 au per year. The code for all the simulations in this paper is written in \textsc{python}. The three-dimensional 
motion of the debris clouds resulting from the collisions is deprojected to the location and geometry of the Fomalhaut 
debris disc. 
Using the expansion velocity of the debris cloud, we also keep track of the optical depth of each cloud.
In Fig.~\ref{fig:fastforward} we show an exaggerated version of the simulation, 
where we enlarged the size of the dusty debris cloud (by giving them a larger expansion velocity) and the optical
depth has an arbitrary value. Furthermore, we fast-forwarded the orbit of the debris clouds in the disc.

For each simulation, we take a collision history of two years. A collision history means that we 
simulate two years of collisions before starting measurements from the simulation. This has been done
to simulate the conditions in the debris disc, because especially the dust clouds from collision between kilometre-sized 
objects can be observed for more than a year before $\tau$ is comparable to the background. 
After the build-up of two years of collision history, we start measuring the optical depth at the position of the background
star. We take time steps of one day and calculate the position of the background star.

For the initial size distribution of the whole disc, which is formed by the collisional cascade, we use the 
mass-loss rate of the Fomalhaut debris disc from 
\citet{/disks/strw1/zeegers/major_project/papers/Acke12_updated}, i.e. $2\times10^{21} \,\mathrm{g}\,\mathrm{yr}^{-1}$. 
We take this mass-loss-rate as a starting point in our simulations.

\begin{figure}
 \begin{center}
 \includegraphics[scale=0.5]{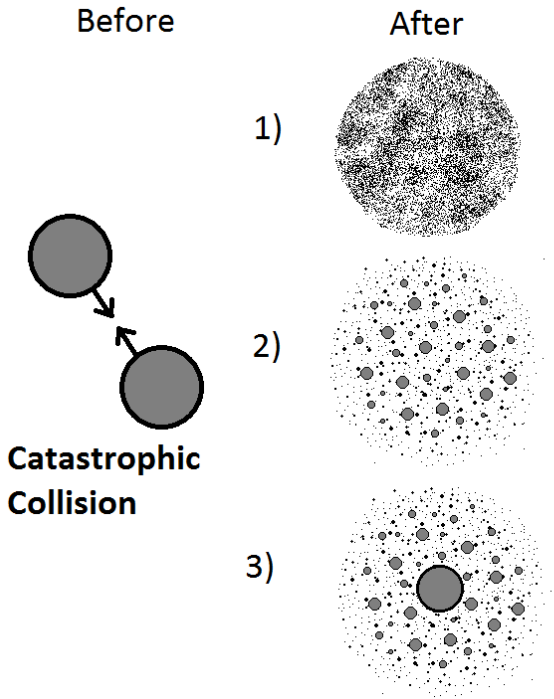}
 \caption{\small{Three considered size distributions of debris after a catastrophic collision. The top figure shows the situation
 where the debris consists only of small micrometer-sized dust particles. The middle figure shows the case where the debris 
 follows a size distribution of a collisional cascade. The bottom figure shows the case where the debris again follows a 
 size distribution but half of the mass is contained in the largest remnant particle. }}
 \label{fig:rubble}
  \end{center}
\end{figure}

We consider the following the size distribution of the debris resulting from 
the collisions.
\begin{enumerate}
\item Total destruction to fine dust.
\item Destruction of the planetesimal with a size distribution following the collisional cascade model.
\item Destruction of the planetesimal with a size distribution following the collisional cascade model 
and with a largest remnant.
\end{enumerate}
\noindent These collision scenarios can be seen in Fig.~\ref{fig:rubble}.  

The first size distribution allows us to explore the extreme upper limit of the amount of dust caused 
by a collision, because it assumes that
the planetesimals will immediately pulverize to dust when they collide. 
When for instance two colliding planetesimals
of 10 km in diameter are ground up to small dust particles (8 $\mu$m) immediately after a collision, 
the resulting dust cloud can expand up 
to $\sim8\times10^{11}\, \mathrm{km}^2$ before the value of $\tau$ drops below 1. If this were the case, we would be
certain that we can detect all the dust clouds that we simulate in this paper using the background star.
This is an extreme case which of course is far from reality, but if the clouds were already unobservable in this
scenario there would be no need for further investigation. During the remainder of the paper we will not consider the fine
dust scenario again.  
The second approach to the debris size is that the debris will follow the size distribution of a collisional cascade.
In our first two scenarios, we will use this distribution, where the largest 
particles will be boulders with a diameter of 10 m and the smallest particles are $1-\mu$m-sized dust particles, 
where particles smaller than 
$8\,\mu$m in diameter are removed instantly by the radiation pressure
of the central star and do not contribute to the expanding orbiting dust clouds.

The third size distribution of debris includes a largest remnant particle. The second largest particle is 
then the largest particle of the collisional cascade and the largest remnant is chosen as half the size of the planetesimal.
This size distribution will be used in the third simulation.

The simulation of the first scenario consists of a 1000 collisions per day in the whole disc between planetesimals of 1 km in diameter. 
In the second and third simulations, we use a distribution of planetesimal sizes, drawn from the collisional 
cascade model. 
We use $q=1.83$ (which is the classical parameter for a self-similar collisional cascade 
\citep{/disks/strw1/zeegers/major_project/papers/Dohnanyi69}) and equation~\ref{eq:collisionalcascade} to estimate the 
scaling factor $K$, and with that determine the size distribution of the particles. 
We copy the strategy of \citet{/disks/strw1/zeegers/major_project/papers/Wyatt02} to keep the 
ejection velocity of the same for all the debris particles, i.e. 
$v_{\mathrm{ej}}=90 \,\mathrm{m}\mathrm{s}{-1}$ (except for the largest remnant particle when included in the simulation). 
Furthermore, we consider all the planetesimals to lie outside the gravity regime and therefore all the debris particles do 
not need to overcome the gravitational energy of the colliding bodies. 
A planetesimal with a diameter of 25 km does not have an ejection velocity that deviates more than 1 $\mathrm{m}\,\mathrm{s}^{-1}$ from 90 $\mathrm{m}\,\mathrm{s}^{-1}$ .

To summarize, the three scenarios are as follows.

\begin{enumerate}
\item Catastrophic collisions between 1-km-sized planetesimals. Debris size distribution follows the collisional cascade model.
\item Catastrophic collisions between planetesimals with a distribution following the collisional cascade model. Debris size distribution follows the collisional cascade model.
\item Catastrophic collisions between planetesimals with a distribution following the collisional cascade model. Debris size distribution follows the collisional cascade model, 
including a largest remnant.
\end{enumerate}

\subsection{Results of the simulations}

\subsubsection{First scenario}

\begin{figure}
 \begin{center}
 \includegraphics[scale=0.45]{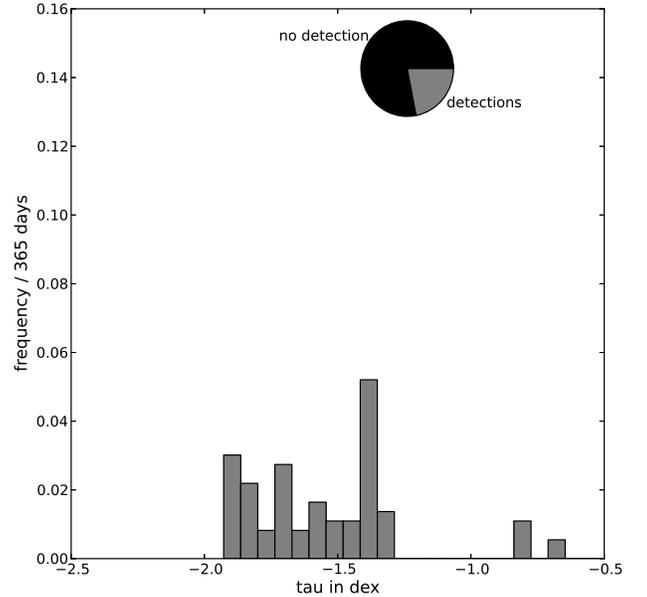}
 \caption{\small{Bar chart showing the probability of the optical depth ($\tau$) above the 
 background value of the optical depth of the Fomalhaut debris disc
 in the simulation of the first scenario. The $x$-axis shows the value of the optical depth given in dex. The $y$-axis shows the frequency
 with which we observe a certain value of $\tau$, i.e the total number of detections of a certain value of $\tau$ divided by
 365 d. The pie chart
 shows the number of detections and non-detections. The modelled optical depth was measured during 365 d. }}
 \label{fig:barchart_fomalhaut}
  \end{center}
\end{figure}

Fig.~\ref{fig:barchart_fomalhaut} shows the result of the simulation of the first scenario 
for which only collisions between 1-km-sized 
planetesimals were considered.
After a simulated year of observations, the detected optical depth has values ranging from -2.5 to -0.5 dex above the 
background value of $\tau=1.2\times10^{-3}$. 
The figure shows the 
frequency of detections per binned value of $\tau$.
For more than $75$ per cent of the time we do not observe a value of $\tau$ above the 
background value, as is shown by the pie chart. 
This is due to the fact that the dust clouds have not expanded so far that they blend in with the 
background. When generating a longer history of collisions than two years, the dense dust
clouds will get the chance to continue expanding and it will become possible to observe dust clouds with a lower $\tau$
value. However, while the clouds expand the value of $\tau$ will drop and it will become harder to measure their 
fluctuations. 

\subsubsection{Second scenario}

\begin{figure}
 \begin{center}
 \includegraphics[scale=0.45]{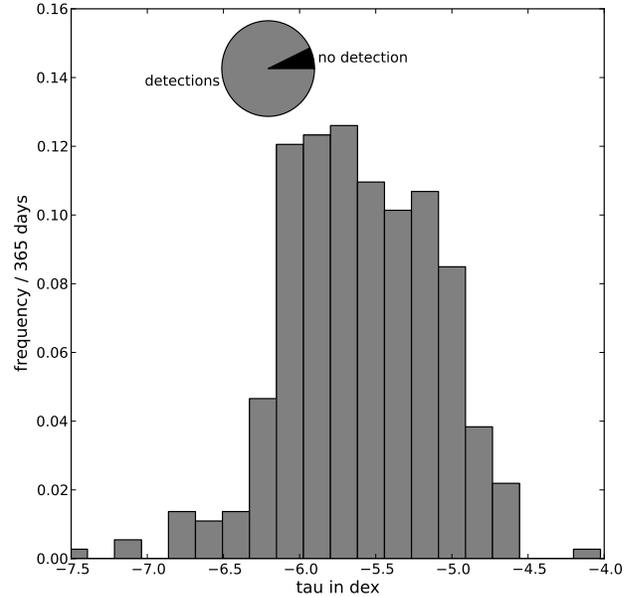}
 \caption{\small{Bar chart showing the probability of the optical depth ($\tau$) above the 
 background value of the optical depth of the Fomalhaut debris disc in the simulation of the second scenario. The pie chart
 shows the amount of observations that we detect an optical depth above the background value. Observations
 were done for 365 d.}}
 \label{fig:barchart_fomalhaut2}
  \end{center}
\end{figure}

Of course it would be very unrealistic only to consider collisions between 1-km-sized planetesimals. Therefore, we introduce
a size distribution for the colliding planetesimals. We draw sizes from a probability distribution by using the collisional 
cascade size distribution as a probability density function. This has been done by scaling equation~\ref{eq:collisionalcascade}
to the total mass in Fomalhaut's debris disc and calculating the resultant factor $K$. 
From this distribution function, we randomly draw planetesimals of a given diameter $D$. We assume that we have two of these
planetesimals of about the same size to create a catastrophic collision, because if the sizes differ too much we 
get a cratering event, which we do not consider in this model. 
\begin{figure}
 \begin{center}
 \advance\leftskip-0.85cm
 \includegraphics[scale=0.5]{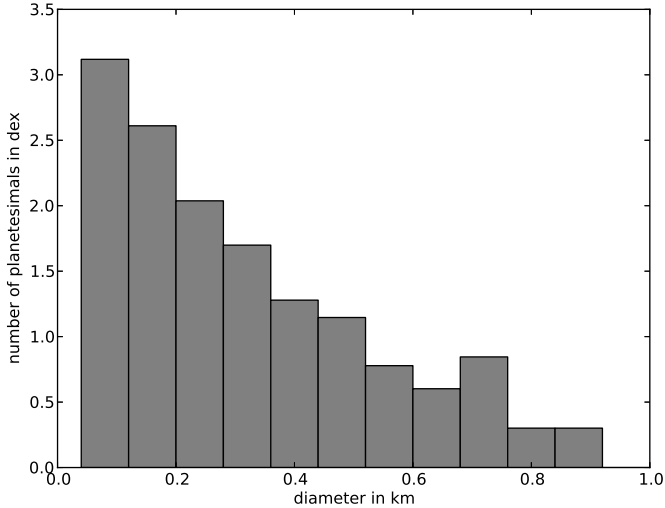}
 \caption{\small{Bar chart showing the distribution of the diameters of the planetesimals for one randomly chosen day in the 
 simulation of the second scenario. The $x$-axis shows the diameter in
 km and the $y$-axis the number of planetesimals in dex. Most of the planetesimals have diameters between 0.1 and 0.2 km.}}
 \label{fig:barchart_1day}
  \end{center}
\end{figure}
The collisional cascade model predicts a large number of small planetesimals. To prevent this
we set a lower limit of $D=100\,\mathrm{ m}$ to the collisional distribution. We also choose an 
upper limit of $D=25 \,\mathrm{ km}$. This upper limit is chosen because above this value it becomes unlikely that there
will be enough planetesimals available to collide with each other over 3 yr. If there were many collisions per day between planetesimals with 
a diameter of 25 km and larger, we would be able to observe these collisions as clumps of reflected light in the disc
as predicted by~\citet{/disks/strw1/zeegers/major_project/papers/Wyatt02} and
these dust clumps are not observed in observations with Herschel 
by~\citet{/disks/strw1/zeegers/major_project/papers/Acke12_updated}. 
Furthermore, it is also the size of the largest comet ever observed in the Solar system ~\citep{2007ApJ...667.1262N}.  
As can be seen in Fig.~\ref{fig:barchart_1day} most of the planetesimal diameters fall between 100 and 200 m. 
The number of 1-km-sized planetesimals is therefore a lot lower than in the previous scenario and we keep track
of more planetesimals. It is not unusual to have more than a 1000 collisions between planetesimals of sizes between 0.1 and 25 km
in an area of $100\,\mathrm{au}^2$ per day.
After generating the collisions, we let the dust clouds expand as in the previous scenario and calculate their 
expansion and Keplerian motion within the disc. We calculate the position of the background star 
and measure the optical depth of 
the collisions during a modelled year. The results can be seen in Fig. ~\ref{fig:barchart_fomalhaut2}.
The values of the optical depth that we measure from this scenario are considerably lower than in the first 
scenario. This difference emanates from the different initial distribution of the particle size. Most of the colliding planetesimals
have diameters between 100 m and 200 m. Although there are more collisions, the resulting debris per collision
is considerably less when compared to the collisions between 1-km-sized planetesimals. This in turn means that it takes less time 
for the debris cloud to blend into the background of the disc.

\subsubsection{Third scenario}

Two colliding planetesimals will not nescessarily completely pulverize into debris consisting of metre-sized boulders up to small dust 
particles. 
There are many other collision scenarios possible in which the largest debris particle will be larger than 10 m size 
boulders.
It is more common that the kinetic energy resulting from the collision is not high enough to pulverize both 
planetesimals. 
We therefore consider another scenario. We assume like \citet{/disks/strw1/zeegers/major_project/papers/Kenyon05}
that half of the mass of 
both planetesimals remains intact in the form of a largest remnant.
When adding a largest remnant to our models the number of collisions needed to produce the same amount 
of dust needed per day doubles, because half of the mass is locked up in the largest remnant. 
The expansion velocity of the dust clouds is again 90 $\mathrm{m}\,\mathrm{s}^{-1}$.
\begin{figure}
 \begin{center}
 \includegraphics[scale=0.45]{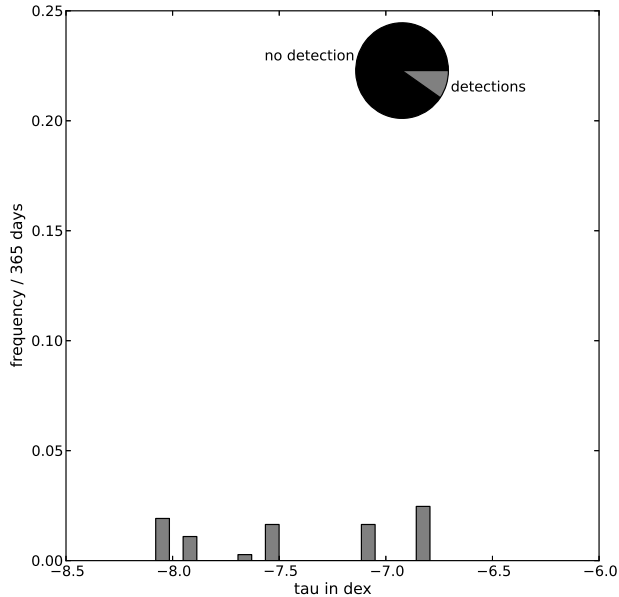}
 \caption{\small{Bar chart showing the probability of detecting the optical depth ($\tau$) above the 
  background value of the optical depth of the Fomalhaut debris disc in the simulation of the third scenario. The optical 
 depth is given in dex. The pie chart
 shows the amount of detections and non-detections. The amount of non-detections is high due to the low amount of dust
 produced in collisions with a largest remnant particle with a mass of half the mass of the original planetesimal.}}
 \label{fig:barchart_fomalhaut_3}
  \end{center}
\end{figure}
We can conclude from Fig.~\ref{fig:barchart_fomalhaut_3} that we observe little change in the optical depth. 
The clouds will be dense enough for only a few weeks due to the fact that half of the mass remains locked up in
the largest remnant and most of the colliding planetesimals have sizes $\sim100\,\mathrm{m}$. Therefore, the background star will not
be able to detect the collisions, though there are $\sim6000$ collisions per day in the selected part of the disc.

\subsection{Frequency of Observations}

\begin{figure}
 \begin{center}
 \includegraphics[scale=0.45]{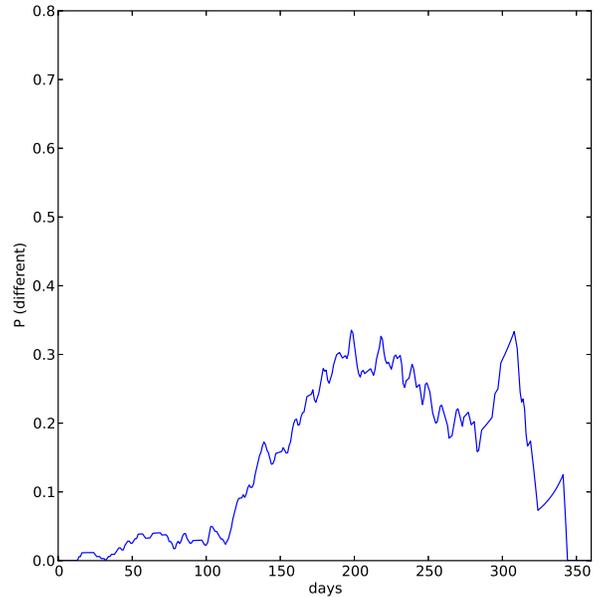}
 \caption{\small{Changes in the optical depth over time in the case of the first scenario. 
 The line indicates a change in the optical depth of at least 1 dex. After a period of 100-150 d there is a 10 per cent chance of observing a change in the 
 optical depth.}}
 \label{fig:frecuency_observing}
  \end{center}
\end{figure}

\begin{figure}
 \begin{center}
 \includegraphics[scale=0.45]{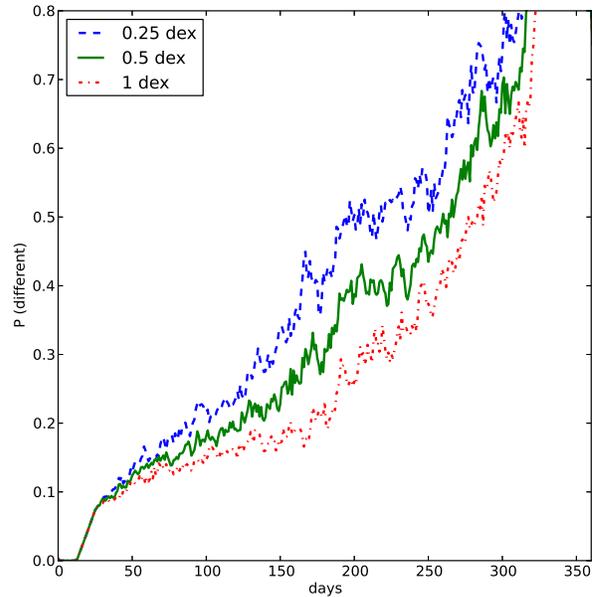}
 \caption{\small{Changes in the optical depth over time for planetesimal sizes ranging between 100 m and 25 km
 (second scenario). 
 The lines indicate changes in the optical depth of 
 0.25 dex 0.5 dex and 1 dex. Similar to Fig.~\ref{fig:frecuency_observing} there is 
 a 10 percent chance of observing a change in the optical depth after 30-50 d indicated by the lines in the legend. }}
 \label{fig:frecuency_observing_100m}
  \end{center}
\end{figure}

\begin{figure}
 \begin{center}
 \includegraphics[scale=0.45]{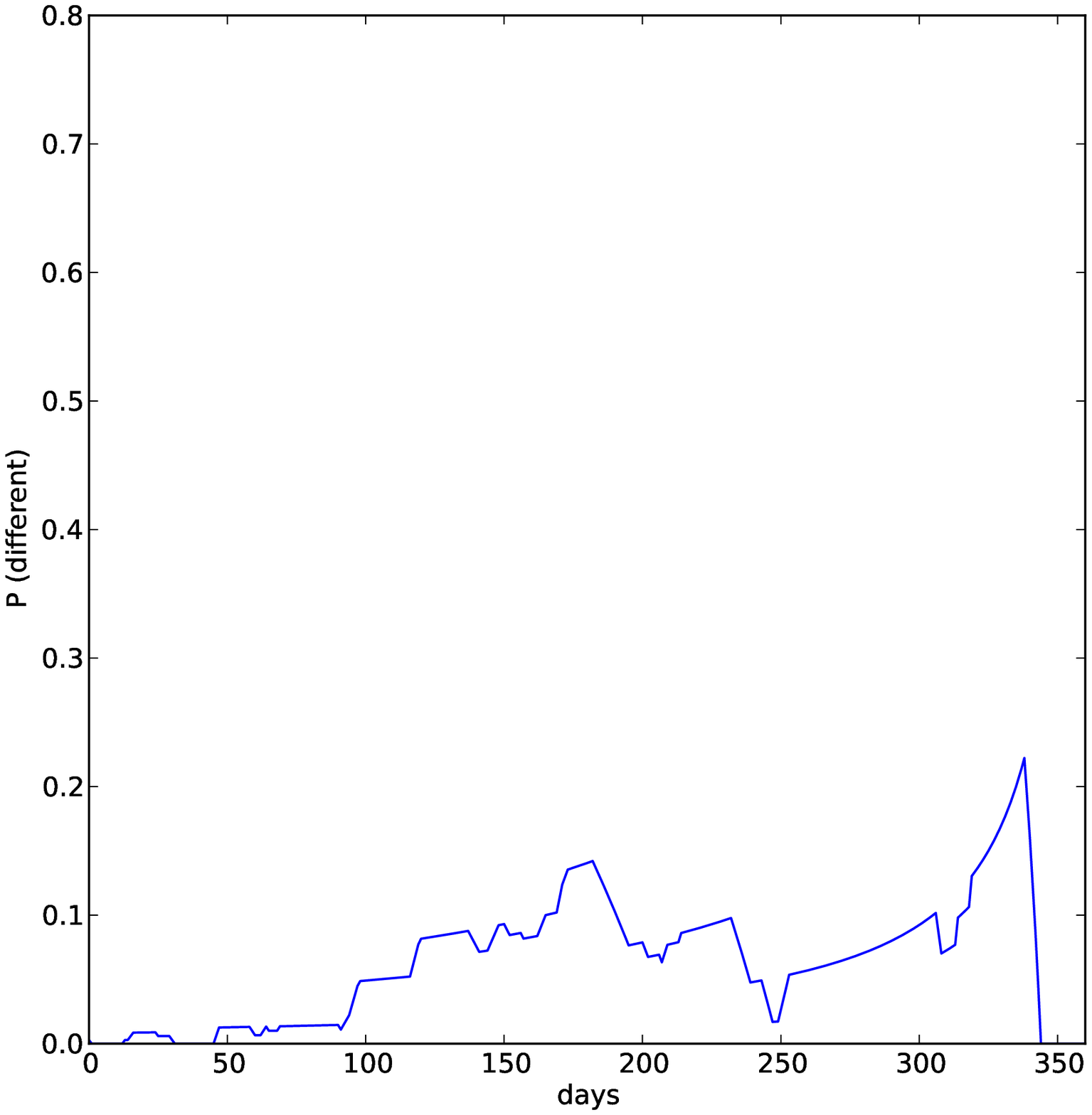}
 \caption{\small{{Changes in the optical depth over time in the third scenario. The line indicates changes in the optical depth of
 0.5 dex. Observations
 were done for 365 d. There are very few detections in this simulation. 
 We do observe a change in the optical depth, but the changes are so small that it will be 
 impossible to detect the dust clouds. In this case we only show the change in optical depth of 0.5 dex, because
 we do not detect larger changes.}}}
 \label{fig:changestau_3}
  \end{center}
\end{figure}

We calculate the simulated value of the optical depth every day, but such frequent observing will not be
necessary. The optical depth will not change dramatically from day to day, because many of the 
detected dust clouds are expand faster than the distance travelled by the star behind the disc in one day. 
We calculate the time between two obervations in which the optical depth will
change the most significant. In Fig.~\ref{fig:frecuency_observing}, the line indicates changes in the optical depth of 
at least 1 dex. From the figure, we can conclude that it is indeed not necessary to observe every day, but
after a period of $\sim150$ d there is a 10 per cent chance that the current observation differs 1 dex from the first 
measurement. In the case of the second scenario (Fig.~\ref{fig:frecuency_observing_100m}), 
this period is $\sim50$ d, due to the higher number of collisions in this scenario. We also plot smaller changes in optical 
depth of 0.25 dex and 0.5 dex.
The same has been done for the third scenario, see Fig.~\ref{fig:changestau_3}. In this case, the chance of
observing a change in the optical depth remains below 0.2 during the whole year due to the lack of detected 
collisions. 

\begin{figure}
 \begin{center}
 \includegraphics[scale=0.45]{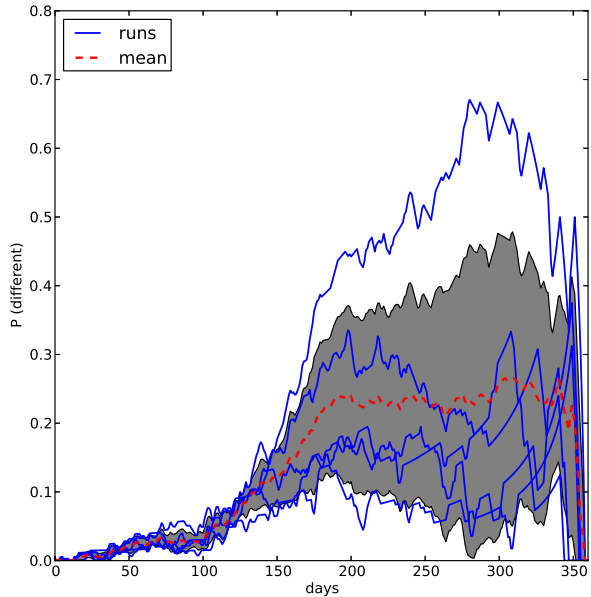}
 \caption{\small{Changes in the optical depth over time in the case of the orginal run of the first scenario
 and 4 other runs with
 equal conditions, but with a different random distribution of the collisions. The blue lines show the changes in 
 the optical depth of 0.5 dex for the 5 runs. The dashed line shows the mean of the five runs and the grey area indicates
 the $1\sigma$ standard deviation.}}
 \label{fig:frecuency_observing_diffmodels}
  \end{center}
\end{figure}

Fig.~\ref{fig:frecuency_observing_diffmodels} shows five different runs of the first scenario. 
The different lines show the changes in the optical depth of 0.5 dex for the five runs. The dashed line indicates
the mean value of the runs.
Due to the number of measurements (365 measurements, 1 every day), we consider the second half of the 
graph less accurate, which is indicated by the $1\sigma$ standard deviation in grey. 
The chance to observe a difference in the optical depth after 365 d for instance,
depends only one one measurement since we only measured for 365 d. 
This effect is also shown in Fig.~\ref{fig:frecuency_observing_diffmodels}. The models are the same as the 
original run of the first scenario (run 1) except for a different random distribution of the collisions. 
Until 150 d, all the runs
behave in the same way. After that period, the lack of measurements explains the difference between the five runs.
The chance to observe a difference in the optical depth after 365 d is zero in most cases, because the optical
depth was equal to the background value on the first day as well as on the $365^{\mathrm{th}}$ day.

\section{Edge-on Discs}

To improve the number of detections and especially the ones involving collisions between large objects ($> 1 \,\mathrm{km}$),
we consider egde-on discs.
An edge-on system allows us to observe both sides of the ring (these systems have an inclination of $\sim90^{\circ}$. 
This method can be for instance applied to the 12 Myr old dwarf star au Microscopii.
As a first approach, we changed the inclination by tilting the original simulation of Fomalhaut to $90^{\circ}$. This means
that we would expect to measure twice as much collisions due to the fact that we are now looking through the front
and the rear part of the ring. 
\begin{figure}
 \begin{center}
 \includegraphics[scale=0.45]{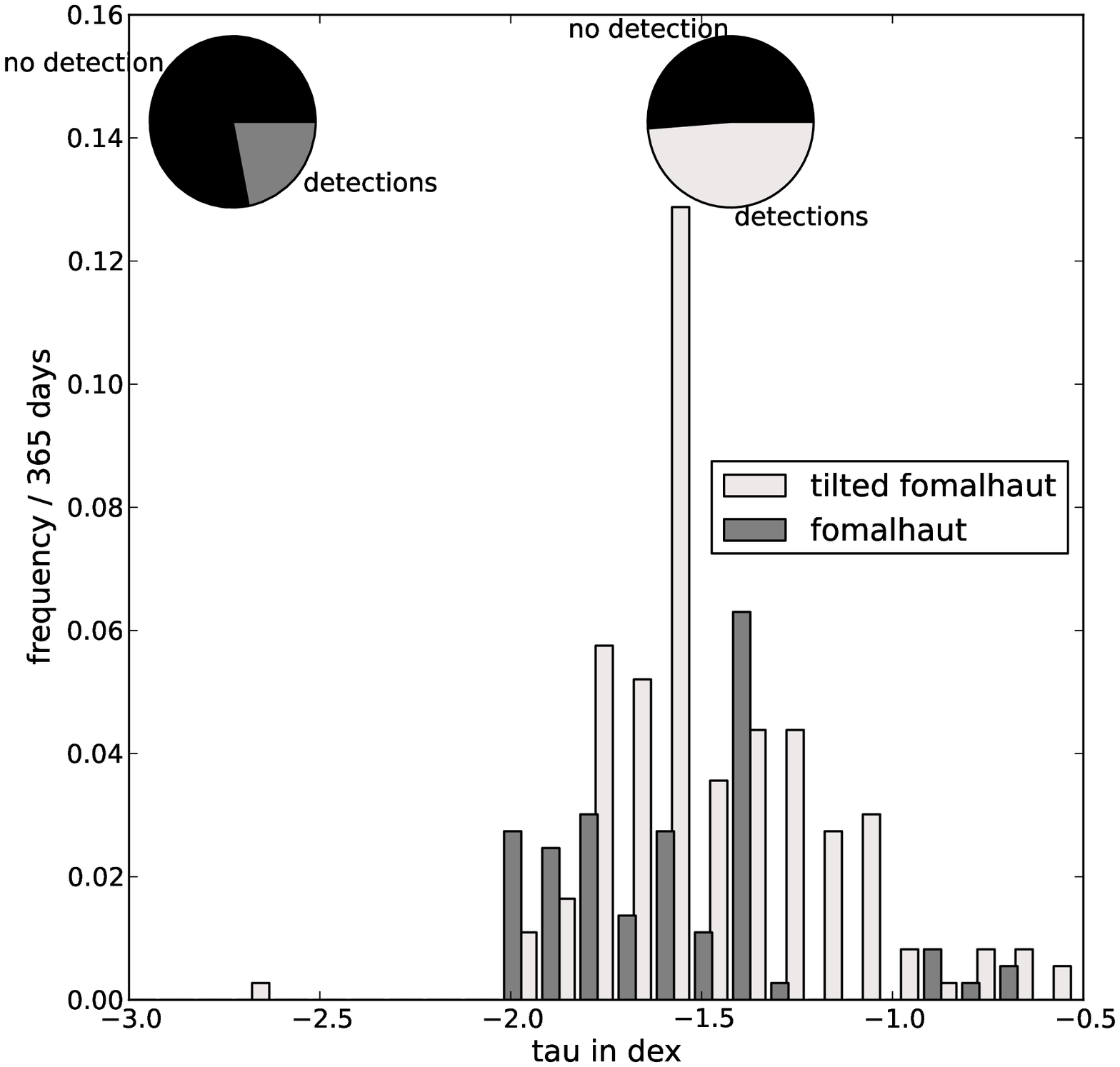}
 \caption{\small{Bar chart showing the probability of detecting $\tau$ for Fomalhaut and a tilted Fomalhaut analog
 ($90^{\circ}$ from face-on). The white bars show the results for the edge-on system and the grey bars show the results for 
 the model of the Fomalhaut debris disc.
 The pie charts show the differences in the number of detections with respect to the inclination. A year of data is simulated.}}
 \label{fig:barchart_sim3}
  \end{center}
\end{figure}
When observing other debris discs, one will encounter a wide variety in inclinations. As shown in Fig.~\ref{fig:barchart_sim3} for this research 
a completely edge-on disc will have the advantage of observing twice as much collision as a face-on system. However,
any inclination between 0 and $90^{\circ}$ will also give an advantage especially when the background star is moving through
one of the ansa.

\section{Other debris discs with transiting background objects}

In this section, we investigate whether there are other debris discs that will move in front of a background object
in the next 5 to 10 yr. 
We studied a selection of nearby debris disc observed at optical wavelengths as shown in Table~\ref{table:alldebrisdisks}. 
These debris discs were selected 
on the basis of their distance and proper motion. 
A small amount of these debris discs will indeed move in front of a background object. These background objects will 
either be background stars or background galaxies. We prefer background stars because they subtend a small area in the 
disc, and these small areas make it possible to study even the smallest dust clouds caused by collisions, although work
has been done on occulting galaxies, see \citet{/disks/strw1/zeegers/major_project/papers/Holwerda07}.

\begin{table*}
 \centering
 \begin{minipage}{140mm}
  \caption{Names of the debris discs and values of the proper motion and distance.}
  \begin{tabular}{l|l l l|l l l|l|l|l l}
  Debris disc&\multicolumn{3}{|c|}{RA (J2000)}&\multicolumn{3}{|c|}{Dec (J2000)}&V mag&Distance in pc&\multicolumn{2}{|c|}{Proper motion $\mathrm{mas}/\mathrm{yr}$}\\
  \cmidrule(r){2-4}
  \cmidrule(r){5-7}
  \cmidrule(r){10-11}
  & h & m & s & \degr & \arcmin & \arcsec & & &RA & Dec \\
  \hline\hline
  Fomalhaut&22&57&39.0465&-29 &37 &20.050&$1.16$&$7.688$&328.95&-164.67\\ \hline
  AU Microscopii&20&45 &09.5318&-31 &20 &27.238&$8.61$&$9.9$&279.96&-360.61\\ \hline
  HD 10647&01&42&29.3157&-53 &44 &27.003&$5.52$&$17$&166.32 &-106.52\\ \hline
  HD 139664&15&41&11.3774&-44 &39 &40.338&$4.64$&$17.52$&-169.17&-266.28 \\ \hline
  HD 53143&06&59&59.6559&-61 &20 &10.255&$6.803$&$18.41$&-161.59&264.67\\ \hline
  Beta Pictoris&05&47&17.088&-51 &03 &59.44&$3.861$&$19.3$&4.65&83.1\\ \hline
  HD 92945&10&43&28.2717&-29 &03 &51.421&$7.719$&$21.6$&-215.23&-50.04\\ \hline
  HD 107146&12 &19 &06.5015&16 &32 &53.869&$7.01$&$28.51$&-174.16&-148.9\\ \hline
  HD 15115&02 &26 &16.2447&06 &17 &33.188&$6.80$&$45$&86.31&-49.97\\ \hline
  HD 15745&02 &32 &55.8103&37 &20 &01.045&$7.49$&$64$&45.82&-47.87\\ \hline
  HD 202628& 21 &18 &27.26879&-43 &20 &04.7450&$6.75$&$24.4$&240.89&21\\ \hline
  \end{tabular}
  \\[2pt]
  \emph{Note}. The data is taken from the SIMBAD astronomical data base~\citep{2000A&AS..143....9W}.  
  \label{table:alldebrisdisks}
\end{minipage}
\end{table*}

\begin{figure*}
 \begin{minipage}{\textwidth}
\begin{center}
 \includegraphics[scale=0.65]{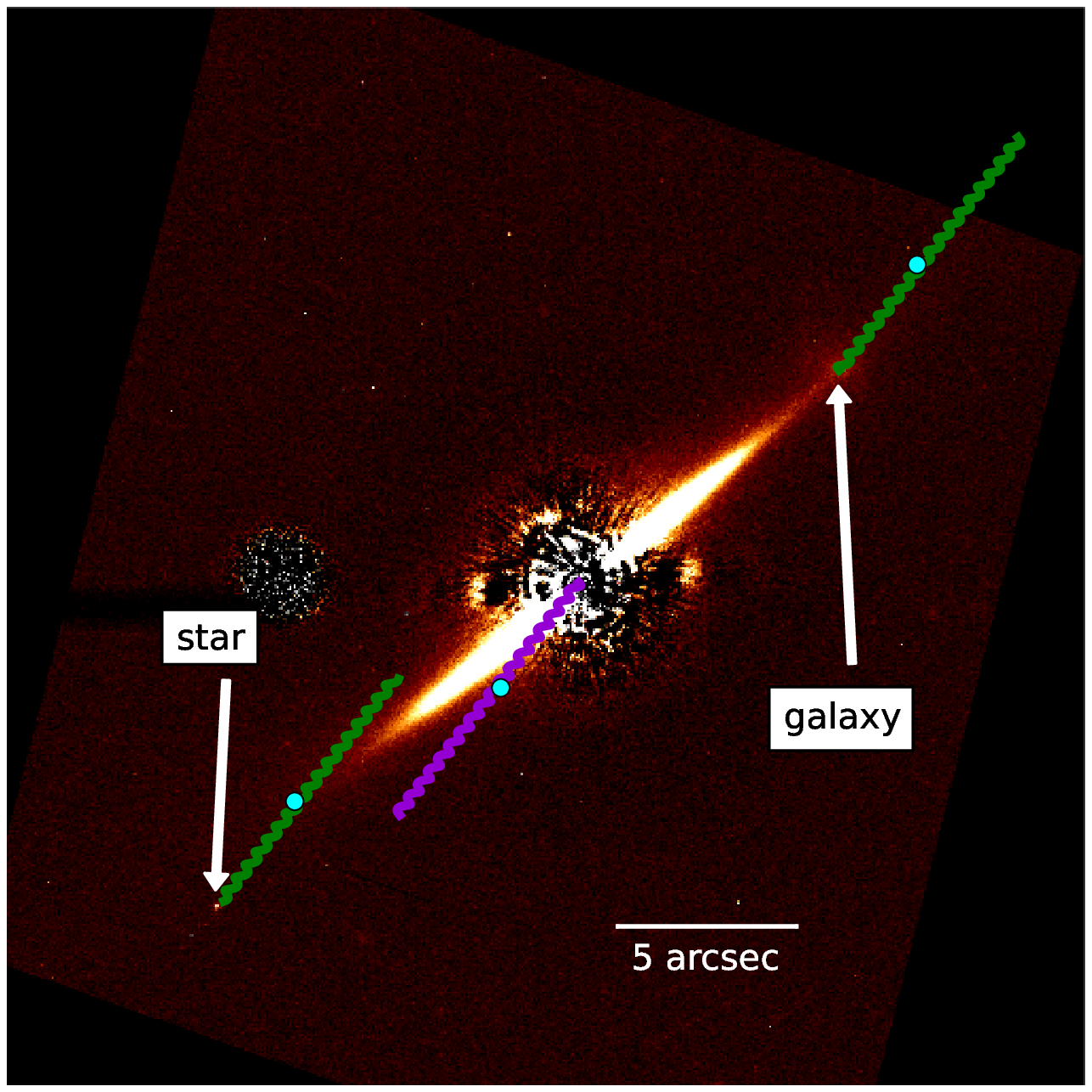}
 \caption{\small{Debris disc of AU Microscopii with background galaxy and star. The combined proper motion and 
 parallax of the debris disc is shown by the purple line. The combined proper motion and 
 parallax of the background star and the background galaxy are shown by the green lines. The blue dots indicate the 
 positions in 2012 October. The background galaxy transited the disk when this image was made in 2004 April. 
 The background star will transit behind the edge of the disc between the beginning of 2013 and 2016.
The image was made using the Advanced Camera for Surveys (ACS) of the \emph{Hubble Space Telescope}.}}
 \label{fig:AUmic_withgalaxy_and_star}
 \includegraphics[scale=0.85]{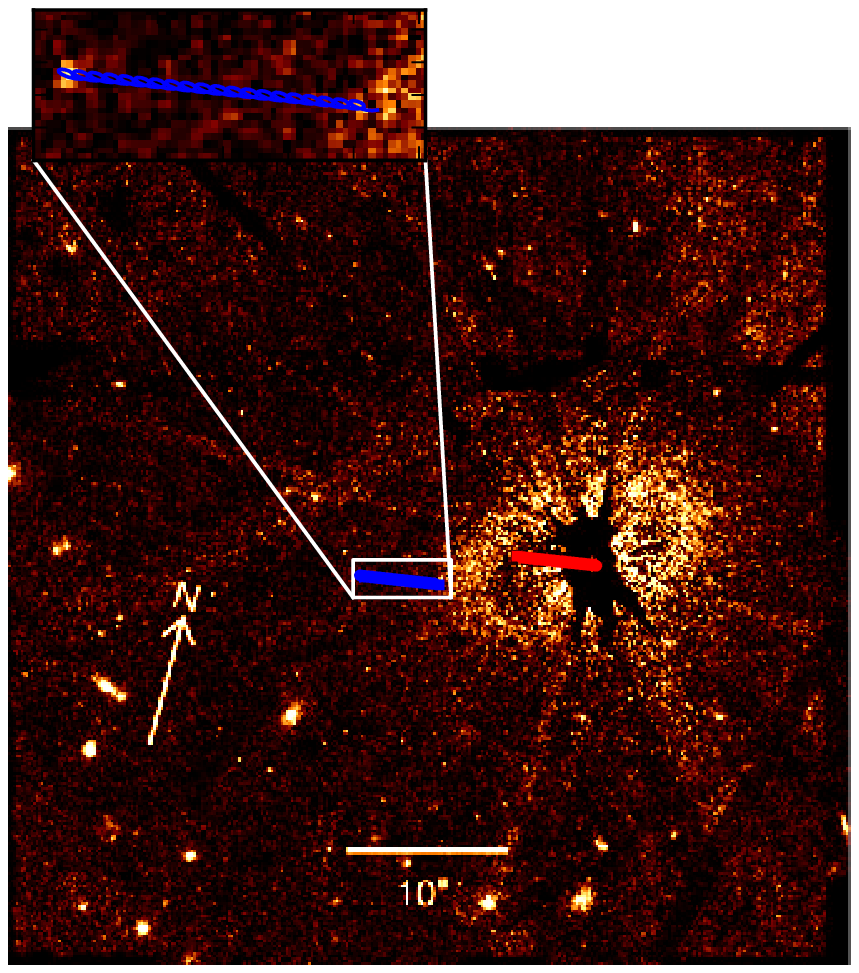}
 \caption{\small{Debris disc of HD 202628. The red line indicates the combined proper motion and parallax of the 
 star and the blue line indicates the motion of the background object relative to the star with debris disc. 
 The object will move behind the debris disc within 20 yr from the 
 moment that the image was taken on the 15th of 2011 May. The brightness profile of the object is more similar to
 that of a galaxy than to that of a star, but further investigation of the object is necessary to determine whether
 the object is a star or a galaxy. The image was made using the Space Telescope Imaging Spectrograph (STIS) instrument 
 from the \emph{Hubble Space Telescope} (\emph{HST})
 \citep{/disks/strw1/zeegers/major_project/papers/Krist12}.}}
 \label{fig:HD 202628}
    \end{center}
    \end{minipage}
\end{figure*}

The selection also only holds for background objects with a high relative proper motion. Stars with a proper motion that is smaller than
50 mas/yr in both directions (RA and Dec)
have a very small chance to occult a background object in the next 5 to 10 yr. Furthermore, this
small change in position will only show the same area in the disc, whereas we are interested in changes in the 
optical depth throughout the disc. 
Stars with a debris disc that have a background object passing behind the disc are listed in Table~\ref{table:alldebrisdisks_withtransit}. 
Most of these debris discs have a galaxy as a transiting background object. Galaxies have an effective diameter
at the debris disc that is much larger than that of the star. In the case of HD 107146, the effective surface area of the galaxy 
amounts to 15 700 $\mathrm{au}^2$. Therefore,
it will not be possible to see individual collisions using a background galaxy moving behind the disc, but it will be 
possible to observe the average optical depth of the debris disc. 
In the case of the debris disc of HD 202628 follow up observations have to determine whether the object is a star or 
a galaxy. This debris disc is comparable to Fomalhaut but the distance to the star is 24.4 pc.  
AU Microscopii shows both a transiting galaxy and a transiting star. The star will transit behind the disc in 2013 and the
galaxy already transited the disc in 2005. 
There is one very faint star in the lower left corner of Fig.~\ref{fig:AUmic_withgalaxy_and_star} 
that might transit the disc. Looking at the green line, we can see that this is indeed the 
case. The motion of the background galaxy is indicated by the second green line. 
The positions in 2012 October are indicated by blue dots.
The observations of AU Microscopii (Fig.~\ref{fig:AUmic_withgalaxy_and_star}), 
HD 107146 and Fomalhaut (Fig.~\ref{fig:fomalhaut_withtransit})
were made by using the \emph{HST} and the ACS \citep{1998SPIE.3356..444K}. HD 202628 was observed using the \emph{HST} with the 
STIS instrument. All images were made using a coronagraph.

\begin{table*}
 \centering
 \begin{minipage}{140mm}
  \caption{Names of the debris discs with a background object.}
  \begin{tabular}{l|c|c|c|c}
  Debris disc&RA (J2000)&Dec (J2000)&Background object&Transit\\
  \hline\hline
  Fomalhaut&03 32 55.8442&-09 27 29.744&background star& transit from 2012 - 2016\\ \hline
  AU Mic&20 45 09.5318&-31 20 27.238&background galaxy& transited disc in 2005\\ \hline
  AU Mic&20 45 09.5318&-31 20 27.238&background star& transit 2013 -$\,\sim$2016\\ \hline
  HD 107146&12 19 06.5015&16 32 53.869&background galaxy& transit from 2012 - 2060\\ \hline
  HD 202628& 21 18 27.26879&-43 20 04.7450&background star or galaxy&transit from 2022 - 2031\\ \hline
  \end{tabular}
  \label{table:alldebrisdisks_withtransit}
\end{minipage}
\end{table*}

\begin{figure}
 \begin{center}
    \advance\leftskip-1.6cm
 \includegraphics[scale=1.0]{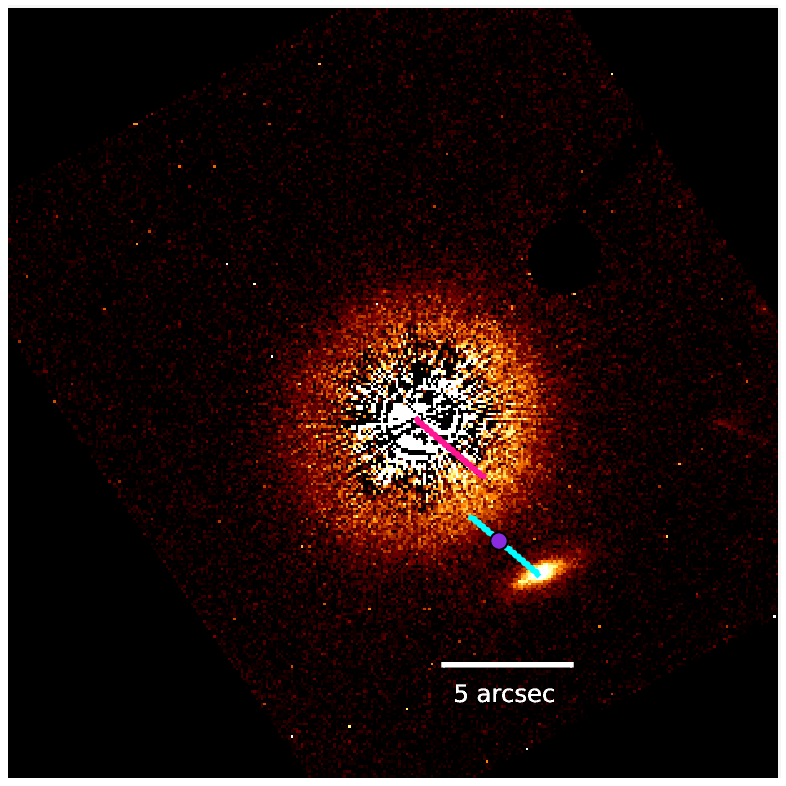}
 \caption{\small{Debris disc around the G2V star HD 107146 with background galaxy. This background galaxy will start moving behind the disc
 within a year and will be totally behind the disc in 2015. The purple dot indicates the position of the galaxy 
 on 2912 September 15. The image was made with the Hubble Space Telescope using the ACS (Advanced Camera
 for Surveys, May 2004).}}
 \label{fig:HD107146_withgalaxy}
  \end{center}
\end{figure}

\section{Summary and Conclusion}
In this paper, we simulate a background star transiting behind the Fomalhaut debris disc. 
This background star gives us an opportunity to observe the distribution of planetesimals as a function of position
in the debris disc
and to test the collisional cascade model, which determines the size distribution of the planetesimals according to 
theoretical models.
We modeled these collisions using three different models that differ in initial size distribution of the 
planetismals and in size distribution of the debris particles. We find that if all the colliding particles are 
pulverized 
1-km-sized planetesimals, the value of the variations of the optical depth due to the dust clumps caused by the collisions 
ranges between $10^{-2.5}-10^{-1.5}$,
establishing a strong upper limit to the experiment.
To make our collision model more realistic, we introduced a size distribution based on the collisional cascade. 
The result of the second scenario shows that the variations of the optical depth decreases to values of $10^{-6.5}-10^{-4.5}$.
The third scenario includes a largest remnant particle. Half the mass of the debris particles remains locked up
in the largest remnant. Again the value of $\tau$ drops, but also the number of detections drops dramatically. These
low mass debris clouds will be impossible to observe.

We should be able to measure changes in the flux from a background star caused by 
catastrophic collisions between large boulders of debris in the disc with diameters of 1 km or larger.  
It is not possible to have the \emph{HST} continuously pointing at Fomalhaut,
but this is not necessary. From 
Fig.~\ref{fig:frecuency_observing} we conclude that
there is a 10 per cent chance that the observed optical depth changes significantly after 100-150 d and from 
Fig.~\ref{fig:frecuency_observing_100m} we conclude that the optical depth changes significantly after 
30-50 d. It would therefore
suffice to observe the debris disc once every 1-2 months.

The values of the optical depth resulting the simulations of our second and third scenario are below the detection limits of 
current telescopes. However, we emphasize that these models are based on the assumption that the size distribution
of the debris resulting from the collision follows a collisional cascade. 
The largest particles in the cascade are 10 meter boulders (scenarios 1 and 2)
or the second largest remnant particle (scenario 3) and the 
smallest particles are 8 $\mu$m particles, just above the blow-out size. The total debris mass is equal to the 
mass-loss rate, which was determined for Fomalhaut by \citet{/disks/strw1/zeegers/major_project/papers/Acke12_updated}. 
They show that this mass-loss rate can be compared to
1000 collisions between 1-km-sized planetesimals, where  
these planetesimals pulverize to small dust particles instead of letting the size distribution of the debris follow a 
collisional cascade. 
Using this particle distribution we underestimate the amount of 
dust generated in the debris disc, but in this way we do not overestimate the number of collisions between
large 1-km-sized planetesimals. This does mean that there are very likely more collisions in the disc than we assume in this
paper, but it is not clear what quantity of dust is produced in cratering collisions and what is produced in 
catastrophic collisions. We emphasize that we used the worst case scenario in this paper and even in these cases 
it is not impossible to detect a change in the optical depth.
Future simulations will therefore have to involve a more realistic collision rate, where the mass-loss rate is built 
up by the amount of blow-out particles that arise from a collision. It is important to  
make some strong assumptions on the contribution of the amount of dust resulting from cratering and from catastrophic
collisions between particles smaller than 100 m.

It is not impossible that we will measure optical depths caused by more than one collision, due to the overlap
of at least two dust clouds resulting from collisions. Our simulations can distinguish between two or more dust clouds, 
but in reality this will 
be harder. The differences in optical depth in the case of overlap is often very large. In the cases we 
encountered, it involved a far expanded dust cloud with a very low value of $\tau$ of the order of $10^{-7}$ and a much
smaller and denser dust cloud. In reality, this would mean that if we observe a change in the optical depth, most of the change
is due to the more compact dust cloud and the signal of the other cloud will be too weak to detect.

Besides the Fomalhaut debris disc, we investigated whether there are more debris discs with a background object
that will transit in the next decade. We
concluded that there are three other debris discs with a background object transiting behind the disc: HD 107146, 
HD 202628 and AU Microscopii. The star HD 107146 has a galaxy that will move behind their discs and AU Microscopii 
has both a star and galaxy transiting behind the disc, but in the case of the star HD 202628
it is not certain whether a galaxy or a star will pass behind the disc. 
We are looking for edge-on systems in particular, because for these systems we can look through two sides of the debris
ring and $\tau$ increases. 

Using a background star to indirectly observe collisions in debris discs is a powerful
method to investigate the size distribution of debris in debris discs.
Though we use a very simple model, it is now possible to put some boundaries to the detectability of the dust 
clouds resulting from the collisions between planetesimals. When observations prove that there is no change in the
brightness of the background star, we can conclude that most of the dust is created in cratering collisions or 
in catastrophic collisions that have a low impact energy whereby at most half of the mass of the planetesimal remains
intact.

Future work includes the role of gas absorption in the line of sight and modeling of other candidate systems. 

\section*{Acknowledgments}
 
We would like to thank Michiel Min,
Mark Wyatt, Karl Stapelfeldt, Mher Kazandjian and Marissa Rosenberg
for their helpful comments. Paul Kalas acknowledges support from NASA NNX11AD21G, NSF AST-0909188 and UCOP LFRP-118057. 

\bibliographystyle{apalike}
\bibliography{zeegers}

\bsp

\end{document}